\documentclass[fleqn,usenatbib]{mnras}

\usepackage{newtxtext,newtxmath}

\usepackage[T1]{fontenc}

\DeclareRobustCommand{\VAN}[3]{#2}
\let\VANthebibliography\thebibliography
\def\thebibliography{\DeclareRobustCommand{\VAN}[3]{##3}\VANthebibliography}

\usepackage{graphicx}	
\usepackage{amsmath}	

\usepackage{xcolor}
\usepackage{wrapfig}

\usepackage{orcidlink}

\newcommand\msun{\ensuremath{\mathrm{M}_{\odot}}}
\newcommand\lsun{\ensuremath{\mathrm{L}_{\odot}}}
\newcommand\kms{km$\,$s$^{-1}$}

\title[Detection of cool molecular gas in REBELS-25]{Direct detection of cool molecular gas in a star-forming galaxy at $z=7.31$}

\author[K. Cescon et al.]{
Karin Cescon,$^{1}$\thanks{E-mail: cescon@strw.leidenuniv.nl} \orcidlink{0000-0003-2160-0454}
Jacqueline A. Hodge,$^{1}$\orcidlink{0000-0001-6586-8845}
Leindert A. Boogaard,$^{1}$\orcidlink{0000-0002-3952-8588}
Hiddo S. B. Algera,$^{2}$\orcidlink{0000-0002-4205-9567}
\and
Lucie E. Rowland,$^{1}$\orcidlink{0009-0009-2671-4160}
Dominik A. Riechers,$^{3}$\orcidlink{0000-0001-9585-1462}
Renske Smit,$^{4}$\orcidlink{0000-0001-8034-7802}
Ilse De Looze,$^{5}$\orcidlink{0000-0001-9419-6355}
Rychard Bouwens,$^{1}$\orcidlink{0000-0002-4989-2471}
\and
Paul van der Werf,$^{1}$\orcidlink{0000-0001-5434-5942}
Manuel Aravena,$^{6, 7}$\orcidlink{0000-0002-6290-3198}
Elisabete da Cunha,$^{8, 9}$\orcidlink{0000-0001-9759-4797}
Pratika Dayal,$^{10, 11, 12}$\orcidlink{0000-0001-8460-1564}
\and
Andrea Ferrara,$^{13}$\orcidlink{0000-0002-9400-7312}
Rebecca Fisher,$^{14}$\orcidlink{0009-0008-3946-0502}
Hanae Inami,$^{15}$\orcidlink{0000-0003-4268-0393}
Pavel E. Mancera Pi\~na,$^{1}$\orcidlink{0000-0001-5175-939X}
\and
Pascal A. Oesch,$^{16, 17}$\orcidlink{0000-0001-5851-6649}
Andrea Pallottini,$^{18}$\orcidlink{0000-0002-7129-5761}
Matus Rybak,$^{1}$\orcidlink{0000-0002-1383-0746}
Sander Schouws,$^{1}$\orcidlink{0000-0001-9746-0924}
\and
Laura Sommovigo,$^{19, 20}$\orcidlink{0000-0002-2906-2200}
Mauro Stefanon$^{21, 22}$\orcidlink{0000-0001-7768-5309}
and Livia Vallini$^{23}$\orcidlink{0000-0002-3258-3672}
\\
$^{1}$Leiden Observatory, Leiden University, PO Box 9513, NL-2300 RA Leiden, The Netherlands\\
$^{2}$Institute of Astronomy and Astrophysics, Academia Sinica, 11F of Astronomy-Mathematics Building, No.1, Sec. 4, Roosevelt Rd, Taipei 106319, Taiwan, R.O.C.\\
$^{3}$Institut f\"ur Astrophysik, Universit\"at zu K\"oln, Z\"ulpicher Stra{\ss}e 77, D-50937 K\"oln, Germany\\
$^{4}$Astrophysics Research Institute, Liverpool John Moores University, 146 Brownlow Hill, Liverpool L3 5RF, UK\\
$^{5}$Sterrenkundig Observatorium, Ghent University, Krĳgslaan 281-S9, B-9000 Ghent, Belgium\\
$^{6}$Instituto de Estudios Astrof\'{\i}sicos, Facultad de Ingenier\'{\i}a y Ciencias, Universidad Diego Portales, Av. Ej\'ercito 441, Santiago, Chile\\
$^{7}$Millenium Nucleus for Galaxies (MINGAL)\\
$^{8}$International Centre for Radio Astronomy Research (ICRAR), The University of Western Australia, M468, 35 Stirling Highway, Crawley, WA 6009, Australia\\
$^{9}$ ARC Center of Excellence for All Sky Astrophysics in 3 Dimensions (ASTRO 3D), Australia\\
$^{10}$Canadian Institute for Theoretical Astrophysics, 60 St George St, University of Toronto, Toronto, ON M5S 3H8, Canada\\
$^{11}$David A. Dunlap Department of Astronomy and Astrophysics, University of Toronto, 50 St George St, Toronto ON M5S 3H4, Canada\\
$^{12}$Department of Physics, 60 St George St, University of Toronto, Toronto, ON M5S 3H8, Canada\\
$^{13}$Scuola Normale Superiore, Piazza dei Cavalieri 7, 56126, Pisa, Italy\\
$^{14}$Jodrell Bank Centre for Astrophysics, Department of Physics and Astronomy, School of Natural Sciences, The University of Manchester, Manchester M13 9PL, UK\\
$^{15}$Hiroshima Astrophysical Science Center, Hiroshima University, 1-3-1 Kagamiyama, Higashi-Hiroshima, Hiroshima 739-8526, Japan\\
$^{16}$Department of Astronomy, University of Geneva, Chemin Pegasi 51, 1290 Versoix, Switzerland\\
$^{17}$Cosmic Dawn Center (DAWN), Niels Bohr Institute, University of Copenhagen, Jagtvej 128, København N, DK-2200, Denmark\\
$^{18}$Dipartimento di Fisica “Enrico Fermi”, Universitá di Pisa, Largo Bruno Pontecorvo 3, Pisa I-56127, Italy\\
$^{19}$Department of Astronomy, Columbia University, 550 W 120th St, New York, NY 10025, USAA\\
$^{20}$Center for Computational Astrophysics, Flatiron Institute, 162 5th Ave, New York, NY 10010, USA\\
$^{21}$Departament d’Astronomia i Astrofísica, Universitat de València, C. Dr. Moliner 50, E-46100 Burjassot, València, Spain\\
$^{22}$Unidad Asociada CSIC “Grupo de Astrofísica Extragaláctica y Cosmología” (Instituto de Física de Cantabria - Universitat de València)\\
$^{23}$INAF, Osservatorio di Astrofisica e Scienza dello Spazio, Via P. Gobetti 93/3, I-40129 Bologna, Italy\\
}

\date{Accepted XXX. Received YYY; in original form ZZZ}

\pubyear{\the\year{}}

\begin{document}
\label{firstpage}
\pagerange{\pageref{firstpage}--\pageref{lastpage}}
\maketitle


\begin{abstract}
We investigate the molecular gas content and interstellar medium (ISM) conditions of REBELS-25, a massive, star-forming galaxy at $z=7.31$. Deep VLA Q-band and ALMA Band 3 observations reveal CO(3--2) and CO(7--6) emission (both at $\sim3.5\sigma$), and provide an upper limit on [C\,\textsc{i}](2--1). From the CMB-corrected CO(3--2) flux--representing the highest-redshift detection of a low-$J$ CO transition to date--we derive a molecular gas mass of $M_{\rm mol}=(1.0\pm0.4)\times10^{11}\,(\alpha_{\rm CO}/(3\,$\msun(K$\,$\kms$\,$pc$^2)^{-1}))$\,\msun, directly confirming the presence of a very massive gas reservoir only $\simeq700\,$Myr after the Big Bang. This implies an extreme gas fraction of $f_{\rm gas}\simeq0.95$, a gas-to-dust ratio of $\delta_{\rm GDR}\simeq6\times10^2$, and a depletion timescale of $\tau_{\rm dep}\simeq1.2\,$Gyr, broadly consistent with extrapolated scaling relations for main-sequence galaxies at lower redshift. Using the radiative transfer code \texttt{TUNER}, we self-consistently model CO and dust continuum emission in the context of the significant CMB background, constraining ISM properties and recovering $M_{\rm mol}= (1.8^{+1.0}_{-0.9})\times10^{11}\,\msun$, independent of assumptions about $r_{31}$ and $\alpha_{\rm CO}$. 
We further discuss the use of alternative molecular gas tracers at early epochs. Combining CO and [C\,\textsc{ii}] measurements, we infer an empirical [C\,\textsc{ii}]-to-H$_2$ conversion factor of $\alpha_{\rm [C\,\textsc{ii}]}=(60\pm25)\,$\msun/\lsun, suggesting [C\,\textsc{ii}] remains a viable molecular gas tracer in the Epoch of Reionization. These results demonstrate the detectability of low-$J$ CO emission even at $z>7$, paving the way for next-generation facilities, and provide critical insights into the rapid mass assembly of galaxies during the first billion years of cosmic history.
\end{abstract}




\section{Introduction}
In recent years, the frontier of galaxy formation and evolution has been pushed deeper into the early Universe, with the detection and characterization of galaxies well within the Epoch of Reionization (EoR; $6 \lesssim z \lesssim 10$) and even beyond. Deep-field surveys with the \textit{Hubble Space Telescope} (HST) and ground-based observatories had already revealed a population of UV-bright galaxies within the EoR, enabling the detection of sources up to $z\simeq11$ \citep[][]{Oesch_2016}, and indicating that vigorous star formation was already underway a few hundred million years after the Big Bang \citep[e.g.,][]{Bouwens_2015, Finkelstein_2016}.

With the launch of the \textit{James Webb Space Telescope} (JWST), a new window on the near- and mid-infrared Universe has opened. JWST campaigns have unveiled a substantial population of galaxies up to $z\gtrsim14$, with some studies even claiming to challenge existing models of early galaxy formation \citep[e.g.,][]{Labbe_2023, Castellano_2024, Carniani_2024, Xiao_2024, Carniani_2025, Schouws_2025, Weibel_2025, Perez_Gozalez_2025}, and in parallel several theoretical works proposing mechanisms that may help reconcile these observations with galaxy formation models, including reduced dust attenuation in young systems, an evolving or top-heavy initial mass function, bursty modes of star formation, and additional energy input associated with accreting black holes \citep[e.g.,][]{Pacucci_2022, Ferrara_2023, Mason_2023, Mauerhofer_2023, Mirocha_2023, Sun_2023, Nikopoulos_2024}.
Moreover, rest-frame optical emission lines are now routinely detected at $z\gtrsim6$, providing robust spectroscopic redshifts and powerful diagnostics of star formation, ionization, and gas-phase metallicity. As previously thought to be the case for only the most massive dusty starbursts at these epochs, these studies reveal the presence of an already chemically mature interstellar medium (ISM) in galaxies observed less than a billion years after the Big Bang \citep[e.g.,][]{Sanders_2025, Pollock_2025, Shapley_2025, Rowland_2026}, implying that galaxy assembly and metal enrichment must have proceeded with remarkable efficiency and speed.
To unravel how these first massive galaxies have formed and evolved, detailed insights into their star formation activity and the gas reservoirs that fuel it are essential.

In the local Universe, spatially resolved studies down to giant molecular cloud (GMC) scales have established that molecular gas is the primary fuel for star formation \citep[e.g.,][]{Leroy_2008, Bigiel_2008, Bigiel_2011, Schinnerer_Leroy_2024}. Measuring the molecular gas content of early galaxies is thus essential for understanding the drivers of their rapid build-up. Molecular hydrogen (H$_2$) constitutes the bulk of this reservoir but, lacking a permanent dipole moment, cannot be directly observed under typical cold ISM conditions.
Instead, molecular gas content is usually derived from proxy tracers, with carbon monoxide ($^{12}$C$^{16}$O, hereafter CO) being the most commonly used. 
Its low rotational level transitions -- CO($J\rightarrow J-1$), with $J \leq3$ -- have low excitation temperatures (few tens of Kelvin degrees) and critical densities ($n_{\rm{crit}}\sim10^3 - 10^4\,$cm$^{-3}$) that make them effective tracers of the of the cold phase of the molecular ISM \citep[][]{Carilli_Walter_2013}. 
Importantly, when only $J>1$ transitions are available, one must infer the ground-state CO(1--0) luminosity, which has been calibrated in the local Universe against the molecular gas mass, from higher-J transitions \citep[e.g.,][]{Bolatto_2013}. Several studies have attempted to constrain the CO spectral line energy distributions (SLEDs) of star-forming galaxies at $z>1$ \citep[SFGs; e.g.,][]{Daddi_2015, Boogaard_2020, Frias_Castillo_2023, Prajapati_2025}, finding significant scatter, especially at $J\geq4$.
These higher-$J$ transitions increasingly trace gas components associated with starburst regions, shocks, or active galactic nuclei (AGN)-driven heating, where the CO ladder shape can vary significantly depending on local ISM conditions and dominant heating mechanisms \citep[e.g.,][]{Carilli_Walter_2013, Rosenberg_2015, Vallini_2018, Vallini_2019, Esposito_2024}.

Therefore, low-$J$ CO observations are particularly valuable for inferring molecular gas masses, especially at high redshift, where ISM conditions and excitation ladders are less constrained.
However, the intrinsic faintness and low excitation temperatures of low-$J$ CO lines make them susceptible to observational limitations that worsen with redshift. The rising temperature of the cosmic microwave background (CMB) -- which increases with redshift as $T_{\rm{CMB}}(z) = T_{\rm{CMB}, z=0}(1+z)$ -- not only contributes to gas heating and excitation, but also reduces the contrast between line emission and background, suppressing observed fluxes \citep[for an overview see][]{da_Cunha_2013}. Combined with the rapid decrease in observed line flux density with redshift due to the increasing luminosity distance, this effect makes detections of low-$J$ CO transitions at $z\gtrsim5$ extremely challenging with current facilities.

Despite these difficulties, low-$J$ CO observations have provided fundamental insights into the molecular gas content of galaxies throughout cosmic time. At intermediate redshift ($0 \lesssim z\lesssim3$), a combination of low-$J$ CO and dust observations has revealed a significant increase in the molecular gas fraction $f_{\rm{gas}} = M_{\rm{mol}}/(M_{\rm{mol}}+M_{\star})$ from a few percent in the local Universe to approximately $50\%$ near the peak of cosmic star formation activity \citep[$z\sim2$;][]{Tacconi_2010, Tacconi_2013, Tacconi_2018, Tacconi_2020}. 
At higher redshift, recent efforts have pushed the limits of current facilities and succeeded in detecting ground-state CO transitions in samples of dusty star-forming galaxies (DSFGs) at $z\sim1-5$ \citep[e.g.,][]{Aravena_2016, Frias_Castillo_2023, Prajapati_2025}. In parallel, large blind spectral surveys such as COLDz and ASPECS have targeted multiple low-$J$ CO transitions over the redshift range $z\sim2-6$ \citep[][]{Walter_2016, Pavesi_2018, Riechers_2019, Decarli_2019}.

Beyond $z\simeq 5$, however, detections of low-$J$ CO emission remain exceedingly rare, and mostly confined to extreme, gas-rich starbursts and quasar hosts \citep[e.g.,][]{Walter_2003, Riechers_2010, Riechers_2013, Strandet_2017, Kaasinen_2024}, which are not representative of the bulk of the galaxy population. Only a handful of more “normal” massive galaxies near the star-forming main sequence (SFMS) have been detected in these transitions so far. \citet{Pavesi_2019} reported CO(2--1) observations in two massive SFGs, including a detection at $z\simeq5.7$, while \citet{Zavala_2022} --exploiting gravitational lensing-- detected CO(2--1) emission in a massive galaxy at $z=6.03$. At even higher redshift, efforts have so far yielded only upper limits, such as in the case of MACS0416\_Y1 at $z=8.31$, which was targeted in CO(2--1) with the VLA, but resulted in a non-detection \citep[][]{Jones_2024a}.
With only a few such measurements available, direct constraints on the cold molecular gas content and excitation of galaxies within the EoR remain almost entirely unexplored.

At cosmic epochs in which CO detections are scarce or infeasible, alternative tracers have been adopted to infer molecular gas content \citep[e.g., for a review see][]{Hodge_da_Cunha_2020}. One widely adopted approach exploits dust continuum emission \citep[e.g.,][]{Scoville_2014, Scoville_2016} which is relatively bright and has now been detected in several tens of galaxies in the EoR up to $z=8.31$ \citep[e.g,][]{Watson_2015, Inami_2022, Tamura_2019}.   
However, this method depends on assumptions about dust temperature ($T_{\rm{dust}}$), opacity, and the gas-to-dust mass ratio ($\delta_{\rm GDR}$), which at high redshift are poorly constrained and are expected to differ significantly from the local Universe \citep[e.g.,][]{Faisst_2020, Sommovigo_2022a}, where these parameters have been calibrated and studied. 

Another potentially promising cold gas tracer is the bright [C\,\textsc{ii}]158$\mu$m fine-structure line (hereafter [C\,\textsc{ii}]). Empirical studies \citep[][]{Zanella_2018, Zhao_2024} have found a relatively tight correlation between the luminosity of this line and the molecular gas mass across a wide range of galaxy types and redshift ($0\lesssim z\lesssim 6$) with an implied [C\,\textsc{ii}]-to-H$_2$ conversion factor of $\alpha_{\rm [C\,\textsc{ii}]}\simeq30\,{\rm M_\odot\,/L_\odot}$.
However, since [C\,\textsc{ii}] arises from multiple ISM phases -- including photodissociation regions (PDRs), ionized gas, diffuse atomic and molecular clouds --, the fraction associated with the cold, star-forming molecular phase depends sensitively on metallicity, density, and the local UV field \citep[e.g.,][]{Vizgan_2022, Gurman_2024, Casavecchia_2025}. 
Furthermore, both observational and theoretical studies report a large scatter in the derived $\alpha_{\rm{[C\,\textsc{ii}]}}$ values \citep[e.g.,][]{Zanella_2018, Madden_2020, Rizzo_2021, Sommovigo_2022a, Sommovigo_2022b, Vallini_2025}, leading to uncertainties of as much as two orders of magnitude in the inferred gas masses. 

Thus, while alternative cold-gas tracers are indispensable when CO observations are infeasible, their use entails significant assumptions and potential biases. 
A robust calibration of these alternative tracers at high redshift remains an essential goal for understanding how to reliably trace the molecular gas content of early galaxies.

In this work, we present new deep VLA observations of CO(3--2) together with ALMA observations targeting CO(7--6) and [C\,\textsc{i}](2--1) in REBELS-25, a massive star-forming galaxy ($M_\star \simeq 2 \times 10^{9}\,{\rm M_\odot}$; ${\rm SFR_{UV+IR}} = 82^{+41}_{-21}\,{\rm M_\odot\,yr^{-1}}$) at $z_{[\text{CII}]}=7.3065\pm0.0001$.  
The detection of multiple CO transitions, combined with extensive ancillary data, enables a direct characterization of the molecular gas reservoir and the physical conditions of the ISM in an evolved galaxy at cosmic dawn.
The paper is organized as follows. Section~\ref{sec:target_and_obs} summarizes the properties of the target and describes the observations and data reduction. Section~\ref{sec:results} presents the analysis and the main results. In Section~\ref{sec:discussion}, we discuss these findings and explore the use of alternative molecular gas tracers, and Section~\ref{sec:conclusions} provides our conclusions.

Throughout this paper, we assume a standard $\Lambda$CDM cosmology with $H_0=67.7\,$km$\,$s$^{-1}\,$Mpc$^{-1}$, $\Omega_m = 0.31$, $\Omega_{\Lambda}=0.69$ \citep[][]{Planck_2020}. This implies a conversion between kiloparsecs and arcseconds of 5.204$\,$pkpc/\arcsec at the redshift of REBELS-25. 

\section{Target and observations}\label{sec:target_and_obs}

\subsection{REBELS-25: a massive star-forming galaxy at $z=7.31$}\label{sec:target}

REBELS-25 (UVISTA–Y–003) is part of the Reionization Era Bright Emission Line Survey (REBELS; \citealt{Bouwens_2022}), an ALMA Large Program that conducted spectral scans of bright far-infrared (FIR) emission lines--[C\,\textsc{ii}]158$\mu$m and [O\,\textsc{iii}]88$\mu$m--in 40 UV-selected SFGs at $z\gtrsim 6.5$. This survey has yielded [C\,\textsc{ii}] and/or dust-continuum detections in 28 of the targeted sources \citep[][Schouws et al., in prep.]{Bouwens_2022, Inami_2022}, dramatically expanding the known population of FIR-bright galaxies in the EoR, and providing a unique opportunity to study ISM and dust properties \citep[][]{Sommovigo_2022a, Ferrara_2022, Dayal_2022, Aravena_2024, Palla_2024}, dust-obscured star formation \citep[][]{Fudamoto_2021, Inami_2022, Algera_2023, Barrufet_2023, van_Leeuwen_2024}, and the assembly histories of massive early galaxies \citep[][]{Topping_2022}.

Within this sample, REBELS-25 stands out as the brightest source in both [C\,\textsc{ii}] and Band 6 dust continuum emission, yet it remains more than an order of magnitude fainter than the extreme dusty starbursts more commonly detected at high redshift.
It is a massive galaxy with log$(M_\ast$/\msun)$=9.30^{+0.12}_{-0.14}$ \citep[][Stefanon et al., in prep.]{Rowland_2026} and a star formation rate of ${\rm SFR_{UV+IR}} = 82^{+41}_{-21}\,{\rm M_\odot\,yr^{-1}}$ at $z_{\rm [CII]}=7.3065\pm0.0001$ \citep[][]{Hygate_2023, Algera_2024b, Fisher_2026}.\footnote{We adopt the SFR$_{\rm IR}$ calculated from the IR luminosity derived from the optically thin MBB fit of \citet{Algera_2024b}, using the conversion of \citet{Inami_2022}, and the SFR$_{\rm UV}$ based on the NIRSpec IFU data \citep[][]{Fisher_2026}.} Its main physical properties are summarized in Table~\ref{tab:R25_properties}.

Multiple ALMA follow-up observations across Bands 3–9 (rest-frame $\sim50-400\,\mu$m) reveal a substantial cold-dust reservoir with $\log(M_{\rm dust}/{\rm M_\odot}) = 8.2^{+0.6}_{-0.4}$, a dust temperature $T_{\rm dust} = 32^{+9}_{-6}\,$K, and a steep emissivity index of $\beta_{\rm IR} = 2.5 \pm 0.4$ \citep[][]{Algera_2024b}. 
These values were derived under the assumption of optically thin emission; however, allowing for moderately optically thick dust does not significantly alter the inferred dust mass or temperature.
 
The presence of a velocity gradient -- initially suggested by the LP [C\,\textsc{ii}] data at a resolution of $\sim1.5\arcsec$ -- was subsequently confirmed by follow-up high-resolution ($\sim0.14\arcsec, 710\,$pc) [C\,\textsc{ii}] observations presented in \citet{Rowland_2024}. These high signal-to-noise ALMA Band 6 data enabled a study of the galaxy’s sub-kiloparsec morphology and kinematics, revealing a surprisingly evolved, dynamically cold rotating disk ($V_{\rm{rot, max}}/\overline{\sigma} = 11^{+6}_{-5}$). 
The presence of such a well-ordered rotating structure is remarkable given the short time available for disk assembly and dynamical settling at $z\simeq7.31$.

REBELS-25 is also included in a JWST NIRSpec/IFU Cycle 1 program (GO-1626, PI: M. Stefanon), which targeted 12 REBELS galaxies in prism mode to probe their rest-frame UV and optical properties \citep[][Stefanon et al., in prep.]{Rowland_2025b, Rowland_2026, Fisher_2025, Fisher_2026, Komarova_2025, algera_2025b, Algera_2026}. These observations confirm the clumpy rest-frame UV morphology previously seen in HST imaging \citep[][]{Stefanon_2019} and, together with the centrally concentrated dust continuum emission, point to a heavily dust-obscured central region (Rowland et al., in prep.). The JWST data further provide a measurement of the gas-phase metallicity of the galaxy \citep[$Z \simeq 0.85\,Z_\odot$;][]{Rowland_2026}.

As already hinted by \citet{Algera_2024a} prior to the JWST observations, this rich data set depicts REBELS-25 as a massive, dynamically evolved, and metal-enriched system $\simeq700\,$Myr after the Big Bang, making it a compelling target for studying the build-up of mass and the efficiency of star formation in the early Universe.

\begin{table}
	\centering
    \caption{Properties of REBELS-25 derived from previous studies. References: [1] Schouws et al. (in prep), [2] \citet{Bouwens_2022}, [3] \citet{Hygate_2023}, [4] \citet{Algera_2024b}, [5] \citet{Inami_2022}, [6] Stefanon et al. (in prep), [7] \citet{Fisher_2026}, [8] \citet{Rowland_2026}. }
	\label{tab:R25_properties}
	\begin{tabular}{lll} 
		\hline
            \hline
		Property & Value & Reference \\
		\hline
		$z_{[\text{CII}]}$ & 7.3065 $\pm$ 0.0001 & [1], [2] \\
            $L_{[\text{CII}]}$ (\lsun) & $1.7\pm0.2\times10^{9}$ & [3] \\
            $L_{\text{IR}}$ (\lsun) & $5.9^{+3.4}_{-1.7}\times10^{11}$ & [4] \\
            $S_{158\,\mu\text{m}}$ ($\mu$Jy) & $260\pm22$ & [5] \\
            log$(M_\ast$/\msun) & $9.30^{+0.12}_{-0.14}$ & [6] \\
            FWHM$_{[\text{CII}]}$ (\kms) & $316\pm15$ & [3] \\
            SFR$_{[\text{CII}]}$ (\msun$\,$yr$^{-1}$) & $246\pm35$ & [3] \\
            SFR$_{\text{IR}}$ (\msun$\,$yr$^{-1}$) & $70^{+41}_{-21}$ & [4], [5] \\
            SFR$_{\text{UV}}$ (\msun$\,$yr$^{-1}$) & $12.0\pm0.5$ & [7] \\
            log$(M_{\rm{dust}}/\rm{M}_\odot)$ & $8.2^{+0.6}_{-0.4}$ & [4] \\
            $T_{\rm{dust}}$ (K) & $32^{+9}_{-6}$ & [4] \\
            $\beta_{\rm{IR}}$ & $2.5\pm0.4$ & [4] \\
            log$(M_{\rm{H_2, [C\,\textsc{ii}]}}/\msun)$ & $10.7\pm0.3$  & [3] \\
            $12+$log(O/H) & $8.62\pm0.17$ & [8] \\
		\hline
	\end{tabular}
\end{table}

\subsection{Observations and data reduction}\label{sec:observations}

\subsubsection{VLA data}\label{subsec:VLAdata}

Observations targeting CO(3--2) emission in REBELS-25 were carried out with the Karl G. Jansky Very Large Array (VLA program ID: 21A-335; PI: J. Hodge) between November 3rd and December 18th 2023 in the Q Band (covering $\simeq40.3-42.3\,$GHz), in the D configuration, which is the most compact one.
The total integration time was around $40\,$h ($\sim20\,$h on source). The data were taken across 15 observation windows under mostly good weather conditions. 
Data calibration, cleaning, and imaging were performed using the CASA software package \citep[Common Astronomy Software Applications for Radio Astronomy;][]{casa_2022}, version 6.5.4.9. We used the data calibrated with version 2023.1.0.124 of the VLA pipeline. Given the low signal-to-noise ratio (S/N) of the line, we also performed a manual calibration by re-running the pipeline steps. We visually inspected the main diagnostic plots (including bandpass and gain solutions), applied additional flagging where appropriate, and compared the resulting calibrated data to the standard pipeline products, finding consistent results.

Prior to analysis, all 15 observations were concatenated, and the relevant target field was split out using CASA routines. Imaging was performed with the \texttt{hogbom} deconvolver, \texttt{standard} gridding, and \texttt{natural} visibility weighting scheme, the latter chosen to optimize sensitivity. We produced both dirty and cleaned images and cubes. The cleaned products were obtained by setting the threshold to \texttt{nsigma}=1.5 as determined from robust statistics,\footnote{\hyperlink{}{https://casadocs.readthedocs.io/en/stable/api/tt/casatasks.imaging.tclean.html}} and applying a circular mask centered on the source.

Using the CASA task \texttt{tclean}, we generated spectral cubes with a channel width of $50\,$\kms. A continuum map was also created by collapsing the full spectral coverage of the observations (observed frequency $40.3-42.3\,$GHz; rest-frame wavelength $7.09-7.43\,$mm).
The resulting synthesized beams for the spectral cubes (median among channels) and the continuum maps are $2.26\arcsec \times 1.86\arcsec$ and $2.27\arcsec \times 1.86\arcsec$, respectively.  
No continuum emission was detected, yielding a $3\sigma$ upper limit of $16\,\mu$Jy$\,$beam$^{-1}$. This non-detection is consistent with the extrapolation to lower frequencies of the modified blackbody model fitted to the multi-band dust continuum data presented in \citet{Algera_2024b}. The best-fit model for this source has a steep emissivity index ($\beta_{\rm{IR}} = 2.5 \pm 0.4$), leading to a low flux density when extrapolating to the Rayleigh–Jeans tail of the dust SED.

\subsubsection{ALMA data}\label{subsec:ALMAdata}

We make use of ALMA Band 3 observations of REBELS-25 (ALMA Project ID: 2021.1.01495.S; PI: I. De Looze), covering the frequency range where the CO(7--6) and [C\,\textsc{i}](2--1) lines are expected based on the [C\,\textsc{ii}] redshift of the source. The observations were carried out between August 2nd and 18th 2022, with a total integration time of $\sim2.9\,$h in configuration C5, with 43 to 45 antennas and precipitable water vapour of 0.8-1.5$\,$mm. The data were calibrated with the ALMA Science Pipeline (version 2021.2.0.128) without further flagging or modification.

As with the VLA data, we concatenated the relevant observations and produced spectral cubes and continuum maps using the same CASA version and cleaning parameters as in Section$\,$\ref{subsec:VLAdata}. Since continuum emission is detected at the position of the source \citep[as already reported in][]{Algera_2024b}, we generated new spectral cubes in which we subtracted the continuum using the CASA task \texttt{uvcontsub} with \texttt{fitorder=0}. Channels within $\pm1.25\times$FWHM$_{\rm{[C\,\textsc{ii}]}}$ around the frequencies expected for CO(7--6) and [C\,\textsc{i}](2--1) were excluded from the fit.
We then re-imaged the continuum map by collapsing the four spectral windows over the observed frequency range $94.6-110.4\,$GHz (rest-frame $327-382\,\mu$m), excluding channels within $\pm2\times$FWHM$_{\rm{[C\,\textsc{ii}]}}$ around the expected CO(7--6) and [C\,\textsc{i}](2--1) frequencies. In this map, continuum emission is detected at the $\gtrsim 4\sigma$ level, consistent with \citet{Algera_2024b}.
From the continuum-subtracted visibilities, we produced spectral cubes with a channel width of $84\,$\kms. All cubes and images were tapered, yielding synthesized beams of $1.00\arcsec \times 0.94\arcsec$ for the spectral cubes (median among channels) and $0.97\arcsec \times 0.90\arcsec$ for the continuum maps.

\begin{figure}
 \includegraphics[width=\columnwidth,trim={0.1cm 0.1cm 0.1cm 0.1cm},clip]{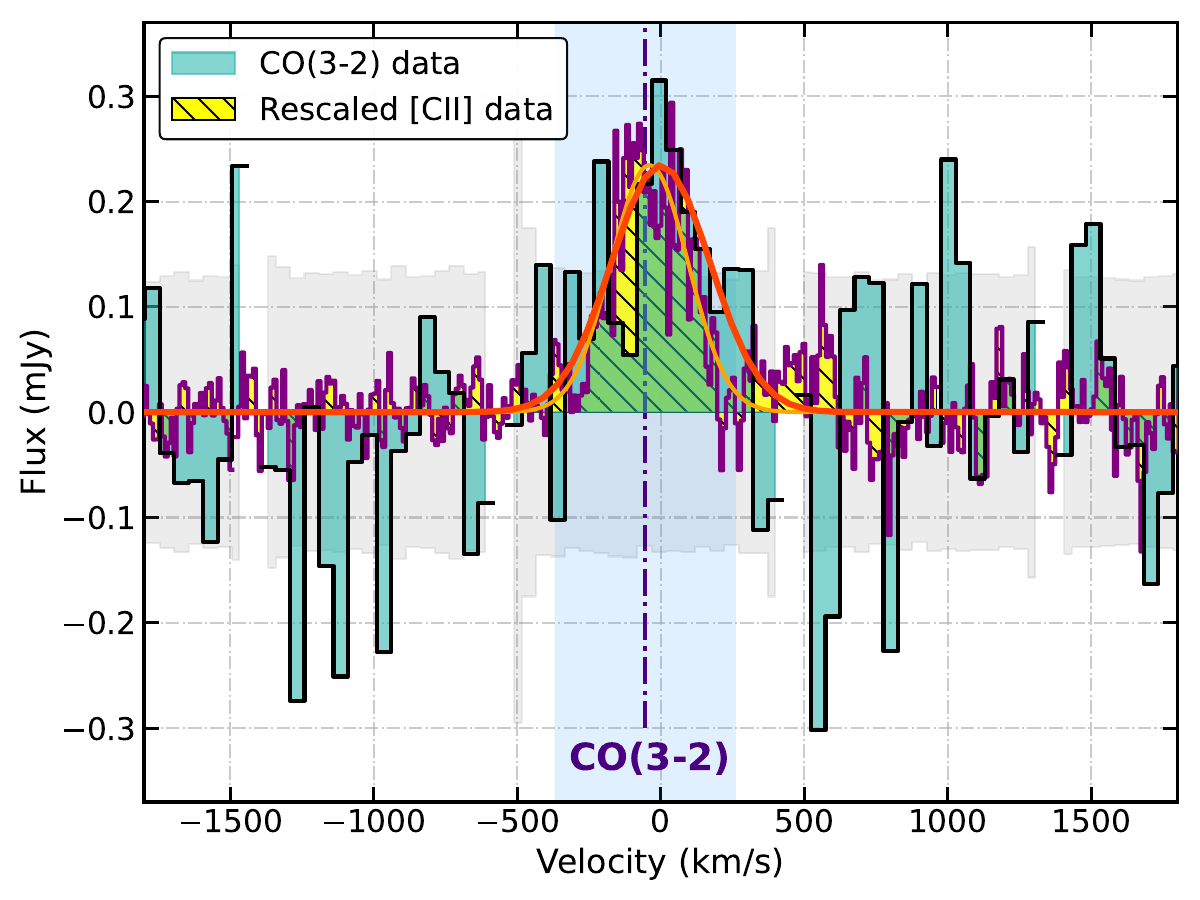}
 \includegraphics[width=\columnwidth,trim={0.1cm 0.4cm 0.1cm 0.1cm},clip]{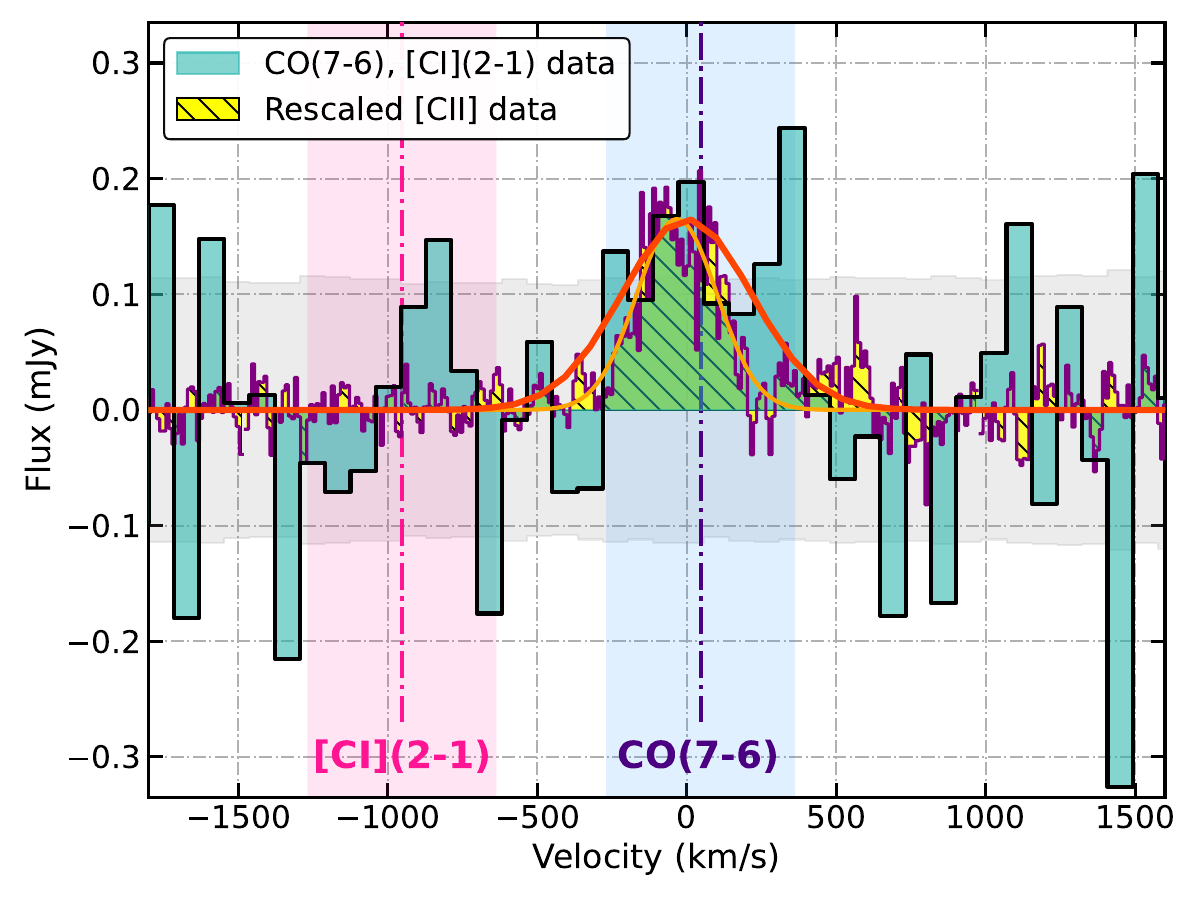}
 \caption{Spectra of REBELS-25 showing the detected line emission of CO(3--2) (\textit{top panel}) and CO(7--6) together with the undetected [C\,\textsc{i}](2--1) emission (\textit{bottom panel}). In each panel, we show the data presented in this paper as an aqua histogram and the respective $\pm 1\sigma$ as a grey shaded area. We also display a comparison with the [C\,\textsc{ii}] spectrum which is detected at high S/N in this source (yellow hatched histogram), rescaled to match the amplitude of the CO emission. Vertical dash-dotted lines and shaded areas show the expected frequency and $\pm 1 \times$FWHM$_{\rm{[C\,\textsc{ii}]}}$ region of each line based on the [C\,\textsc{ii}]-based redshift of the source. Red lines show the best-fit Gaussian to the CO line when leaving all parameters free, while orange lines show the best-fit to the [C\,\textsc{ii}] data rescaled to the amplitude of the CO emission. In the top panel, there are some gaps in the CO(3--2) data due to fully flagged channels at the edge of spectral windows.}
 \label{fig:spectrum}
\end{figure} 

\section{Analysis and results}\label{sec:results}

We produced velocity-integrated intensity (moment 0) maps of the line emission using the \texttt{tclean} task in \texttt{CASA}, adopting multi-frequency synthesis (\texttt{specmode=`mfs'}) and including the emission within $\pm1\times$FWHM$_{\rm{[C\,\textsc{ii}]}}$ of the frequency expected for the emission lines from the [C\,\textsc{ii}] redshift (blue shaded regions in Fig.~\ref{fig:spectrum}).

\begin{figure*}
 \includegraphics[width=\columnwidth,trim={0.1cm 0.4cm 0.1cm 0.1cm},clip]{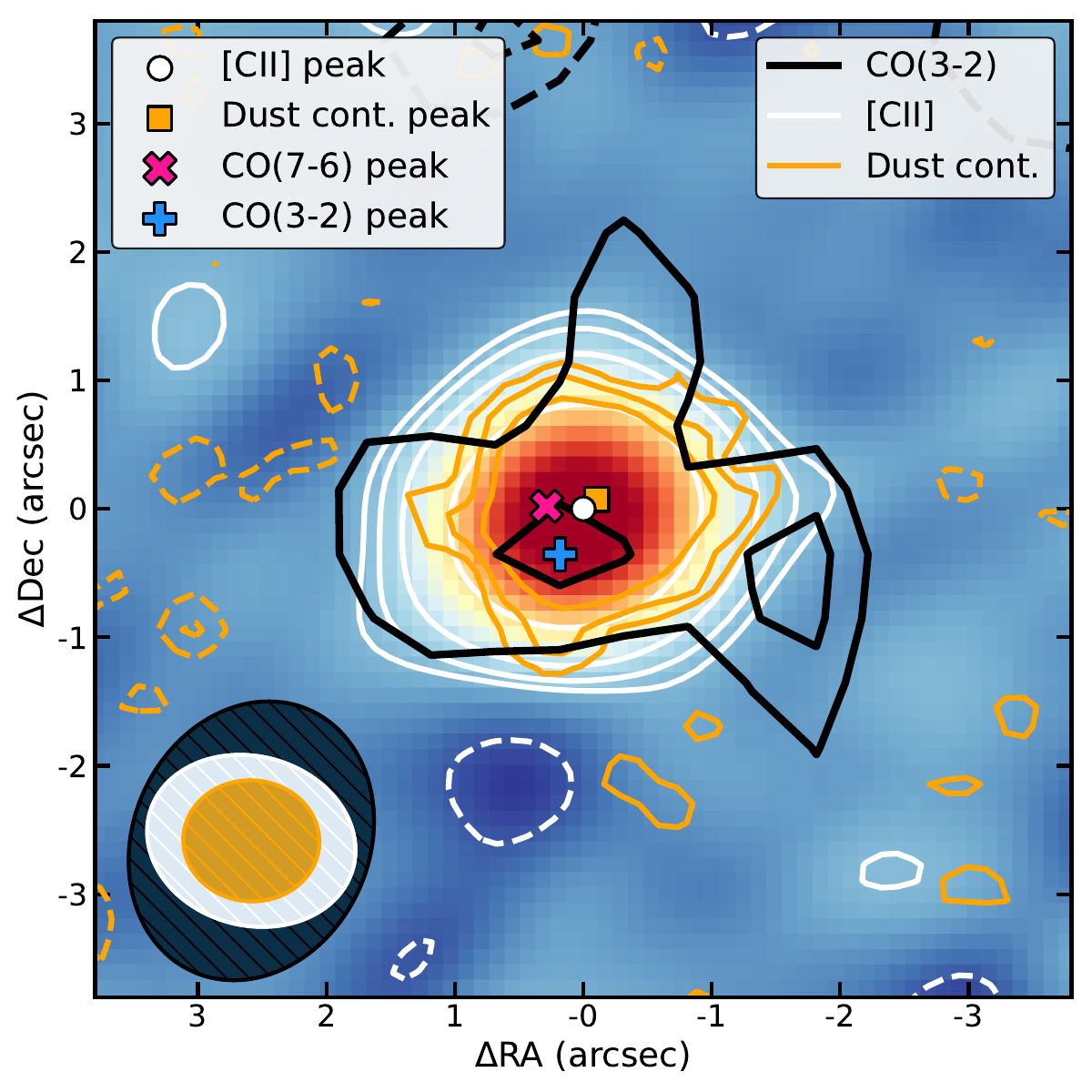}
 \includegraphics[width=\columnwidth,trim={0.1cm 0.4cm 0.1cm 0.1cm},clip]{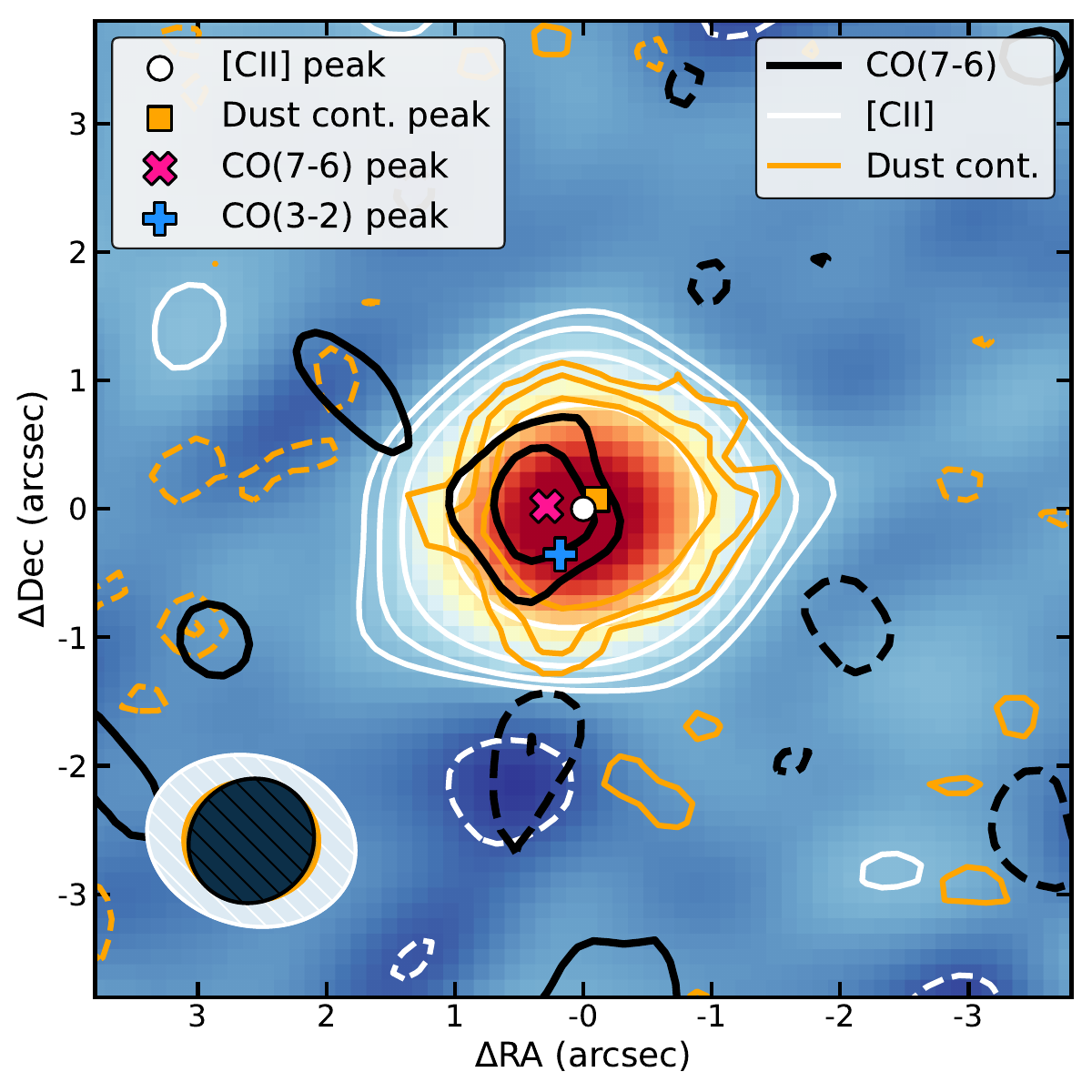}
 \caption{{Black contours ($[-3,-2,2,3]\sigma$) show the velocity-integrated emission of CO(3--2) (\textit{left panel}) and of CO(7--6) (\textit{right panel}) overlaid on the [C\,\textsc{ii}] moment 0 map obtained from the REBELS large program data. We also show the contours for the [C\,\textsc{ii}] emission (white, $[-3,-2,2,3,5,12]\sigma$) and ALMA band 6 continuum tracing the dust emission (orange, $[-3,-2,2,3,5]\sigma$). Synthesized beams for each dataset are shown as ellipses of the corresponding colour in the bottom left corner of the plot. The positions of the peak of the detected CO lines are consistent with the known position of the source.}}
 \label{fig:mom0_separated}
\end{figure*}

In the spectra (Fig.~\ref{fig:spectrum}; see Sec.~\ref{line_luminosities} for details) and in the moment-0 maps, we detect emission features in both datasets. Fig.~\ref{fig:mom0_separated} shows the velocity-integrated CO(3--2) (\textit{left panel}) and CO(7--6) (\textit{right panel}) emission, overlaid on the [C\,\textsc{ii}] moment-0 map from the REBELS large program data, which has a spatial resolution comparable to that of the data presented in this work.
CO(3--2) is detected at a $3.4\sigma$ significance in the VLA data, while CO(7--6) is detected at $3.5\sigma$ in the ALMA data. [C\,\textsc{i}](2--1) emission remains instead undetected. The detected CO emission is spatially coincident with both the [C\,\textsc{ii}] and the ALMA Band 6 continuum emission tracing the dust.

\subsection{Observed line luminosities}\label{line_luminosities}

Due to the limited S/N, we extracted spectra from the dirty cubes.
For the VLA data, the size of the synthesized beam is $\sim2.3$\arcsec, corresponding to $\sim12\,$kpc at the redshift of the source. The emission shows two peaks at $\gtrsim3\sigma$ close to the position of the source. We therefore extracted the spectrum from an aperture centered on the morphological center of the [C\,\textsc{ii}] emission, determined by \citet{Rowland_2024} from fitting the high-resolution data using \texttt{CANNUBI},\footnote{\hyperlink{}{https://www.filippofraternali.com/cannubi}} a forward-modelling software that finds the geometrical parameters of emission-line observations taking into account beam effects \citep[][]{Mancera_Pina_2020, Fraternali_2021, Mancera_Pina_2022, Roman-Oliveira_2023}. Extracting instead the spectrum from the position of the peak of the CO(3--2) emission yields consistent results within the uncertainties.

The native resolution of the ALMA data is higher, and -- considering the limited S/N of the data -- we tapered the visibilities to $\sim1.0$\arcsec (see Sec.~\ref{subsec:ALMAdata}).
For this dataset, the spectrum was extracted from an aperture centered at the position of the CO(7--6) peak, which is consistent (within the reported ALMA positional accuracy\footnote{ALMA's positional accuracy is roughly given by beam/SNR/0.9 (\hyperlink{}{https://almascience.nrao.edu/proposing/technical-handbook}).}) with the morphological center of the [C\,\textsc{ii}] emission.

The spectra were extracted using circular apertures with radii of $2.5$\arcsec for the VLA data and $1.0$\arcsec for the ALMA data. 
We then fitted a Gaussian function to the spectra around the frequency of the CO lines, as expected from the [C\,\textsc{ii}] redshift. Allowing all parameters to vary, the best-fit Gaussian to the CO(3--2) line yields a peak frequency that corresponds to a redshift of $7.3076 \pm0.0016$ and a FWHM$_{\rm{CO(3-2)}} = 405\pm134\,$\kms, both consistent with the [C\,\textsc{ii}] emission within uncertainties. Owing to the large uncertainty on the FWHM, we repeated the fit with the FWHM fixed to the [C\,\textsc{ii}] value \citep[$316\pm15\,$\kms;][]{Hygate_2023}, obtaining a velocity-integrated flux of $88\pm21\,$mJy$\,$\kms, which we adopt in the following analysis.
This value is consistent with the integrated flux obtained when leaving the FWHM free in the fit ($101 \pm 44\,$mJy$\,$\kms).

We repeated the same procedure with the ALMA data using the continuum-subtracted visibilities. 
For the CO(7--6) line, the frequency of the peak of the best-fit Gaussian function to the data corresponds to a redshift of $7.3074\pm 0.0027$, consistent with the [C\,\textsc{ii}] emission. The FWHM$_{\rm{CO(7-6)}} = 512\pm 231\,$\kms$\,$is less constrained but similar to the one of CO(3--2). Therefore, also in this case we fix the FWHM of the fitting Gaussian function to the [C\,\textsc{ii}] value, obtaining a velocity-integrated flux of $64 \pm 23\,$mJy$\,$\kms, which will be adopted in the following analysis. When leaving the FWHM free in the fit, we derive an integrated flux of $89 \pm 54\,$mJy$\,$\kms.

We converted the measured fluxes into line luminosities using Equation 3 from \citet{Solomon_2005}:
\begin{equation}\label{eq_solomon_line}
    L_{\rm{CO}}^{'} = 3.25 \times 10^{7} S_{\rm{CO}} \Delta v \, \nu_{\rm{obs}}^{-2} (1+z)^{-3} D_{\rm{L}}^2,
\end{equation}
where the CO line luminosity ($L_{\rm{CO}}^{'}$) is measured in K$\,$\kms$\,$pc$^2$, the velocity-integrated flux ($S_{\rm{CO}} \Delta v$) is expressed in Jy$\,$\kms, the observed frequency corresponding to the peak of the emission ($\nu_{\rm{obs}}$) is measured in GHz, and the luminosity distance ($D_{\rm{L}}$) is in Mpc.
Applying this relation, we find $L_{\rm{CO(3-2), obs}}^{'} = (1.6\pm0.4)\times10^{10}\,$K$\,$\kms$\,$pc$^2$ and $L_{\rm{CO(7-6)}, obs}^{'} = (2.1\pm0.8)\times10^{9}\,$K$\,$\kms$\,$pc$^2$.\footnote{We adopt the subscript $_{\rm obs}$ to distinguish the observed line luminosities from the intrinsic ones, which in the following we recover by applying a correction for the effects of the CMB.} For the undetected [C\,\textsc{i}](2--1) emission, we report a $3\sigma$ upper limit of $L_{\rm{[C\,\textsc{i}](2-1), obs}}^{'} < 1.6 \times10^{9}\,$K$\,$\kms$\,$pc$^2$, computed from the moment 0 map as three times the rms within $\pm 1\times\rm{FWHM}_{[\rm{CII}]}$ around the expected frequency, and then converted to line luminosity using Eq.$\,$\ref{eq_solomon_line}.

\subsection{Assessing the fidelity of the line detections}\label{line_fidelity}

To assess the reliability of the CO line detections in REBELS-25, we estimated the probability that a random noise fluctuation in the continuum-subtracted datacubes could mimic a spectral-line signal with properties comparable to those expected for REBELS-25.

We first generated suites of moment-0 maps by integrating the cubes over velocity ranges spanning $\sim (0.5-2)\times$FWHM$_{\rm [C\,\textsc{ii}]}$ and central velocities within $\sim \pm 150\,$\kms of the [C\,\textsc{ii}] redshift. For the VLA datacube, this resulted in 77 moment-0 maps covering central velocities within $\pm151\,$\kms of the [C\,\textsc{ii}] redshift, and integration widths of $151-665\,$\kms. For the ALMA data, owing to the larger channel widths of the cube, 25 moment-0 maps were created, spanning central velocities within $\pm169\,$\kms of the [C\,\textsc{ii}] redshift, and widths of $169-675\,$\kms. 

We then used the Python Blob Detector and Source Finder\footnote{\hyperlink{}{https://pybdsf.readthedocs.io/en/latest/}{https://pybdsf.readthedocs.io}} (\texttt{PyBDSF}; \citealt{Mohan_2015}) to identify all (real or spurious) peaks in the datacubes.
We adopted the same \texttt{PyBDSF} configuration used in the REBELS large program dust-continuum analysis \citep[][]{Inami_2022}, namely a detection threshold of (\textit{peak}, \textit{island}) = (3.3, 2.0), and verified that both CO lines from REBELS-25 are detected with these parameters.\footnote{We define here a detection as a feature identified by \texttt{PyBDSF} to be above the set threshold and within a radius equal to one synthesized beam from the [C\,\textsc{ii}] morphological center of REBELS-25 (corresponding to $\simeq 2\arcsec$ for the VLA cube and $\simeq 1\arcsec$ for the ALMA cube).}
Running \texttt{PyBDSF} on all resulting primary-beam–uncorrected, velocity-integrated maps over the full imaged field of view of each dataset (down to the primary-beam cutoff adopted during imaging, which corresponds to the default cutoff of 0.2), we identified all potential candidate features exceeding the adopted threshold. Features located within one beamsize of the REBELS-25 position were removed, as were duplicate detections lying within one beamsize of each other across the different maps. This procedure yielded a catalogue of 52 unique noise candidates for the VLA datacube and 226 unique candidates for the ALMA cube.

For each of these candidates, we extracted a spectrum at the corresponding spatial position and fitted a one-dimensional Gaussian profile to measure amplitude, centroid frequency, and FWHM. We retained only those spurious features whose fitted properties resemble those expected for a genuine CO line: namely, velocity offsets satisfying $|v_{\rm offset}| < 160\,$\kms (relative to the frequency expected for the CO line as inferred from the [C\,\textsc{ii}] redshift) and linewidths in the range $160 < \mathrm{FWHM} [$\kms$] < 640$. This filtering yielded a subset of “believable” features whose distribution in the $(v_{\rm offset}, {\rm FWHM})$ parameter space overlaps with that in which a real CO line would be expected.
We note that the best-fit Gaussians to the observed emission features peak at velocity offsets with respect to the [C\,\textsc{ii}] redshift of $54$ and $47\,$\kms for the CO(3--2) and CO(7--6) transitions, respectively. These offsets are significantly smaller than the velocity offset threshold of $160\,$\kms adopted in this analysis.

We then estimated the surface density of such believable peaks across the full moment-0 field of view and scaled this density to the effective search area defined by a single synthesized beam at the position of REBELS-25. This approach is quite conservative, since the centroids of the emission features we detect lie well within one beam size of the [C\,\textsc{ii}] position. This procedure provides a direct estimate of the probability that a noise fluctuation with realistic linewidth would appear at the source location and at the expected frequency, and therefore could be misidentified as a CO detection.
Applying this procedure, we find that the probability that the observed CO(3--2) emission is spurious is 1.5\%, while the corresponding probability for the CO(7--6) is 2.5\%. Assuming then that the two probabilities are independent, we can also estimate the probability that at least one of the two detections is spurious as $1-(1-0.015) \times (1-0.025) = 4\%$. This result supports the interpretation that the detected features correspond to CO emission lines from REBELS-25.

\subsection{Estimating a molecular gas mass}\label{molecular_gas}

\subsubsection{The Effect of The Cosmic Microwave Background}\label{CMB_effect}

At the redshift of REBELS-25 the temperature of the CMB is $T_{\rm{CMB}}(z=7.31)\simeq23\,$K, close to the value of the dust temperature measured for this source \citep[$T_{\rm{dust}} = 32^{+9}_{-6}\,$K;][]{Algera_2024b} and to the upper level temperature of the lower $J$ rotational transitions of CO \citep[for $J=3\rightarrow2$, $T_{\rm{ex}}\simeq 33\,$K;][]{Carilli_Walter_2013}. The presence of a warm CMB has a significant effect on the CO luminosity that can be observed and has to be taken into account \citep[for a comprehensive treatment see][]{da_Cunha_2013}. It causes two competing effects: it acts as a background against which lines have to be detected, and it boosts line luminosity due to an increase in the energy density that can excite the emitting gas.
The combined effect is that the observable line brightness is typically reduced relative to the intrinsic brightness, and it is particularly significant for low-excitation lines. 

To recover the intrinsic line luminosities, we estimate a correction factor following \citet{da_Cunha_2013}, so that the intrinsic line luminosity is $L_{\rm{CO}(\textit{J} - \textit{J}-1)}^{'} = L_{\rm{CO}(\textit{J} - \textit{J}-1), obs}^{'} / f_{J \rightarrow J-1}$. 
We assume local thermal equilibrium (LTE), such that the excitation temperature of the line is equal to the gas kinetic temperature. It is physically plausible for the cold gas kinetic temperature to be comparable to, or modestly warmer than, the dust temperature. Hence, we assume $T_{\rm{gas}} \simeq 40\,$K ($\gtrsim T_{\rm{dust}}$).  
Under these assumptions (to which we return in Sec.~\ref{TUNER}), the correction factors are $f_{3\rightarrow2}\sim0.55$ and $f_{7\rightarrow6}\sim0.65$ for the CO(3--2) and CO(7--6) lines, respectively. 
Applying these corrections to the observed line luminosities, we recover $L_{\rm{CO(3-2)}}^{'} = (2.9\pm0.7)\times10^{10}\,$K$\,$\kms$\,$pc$^2$ and $L_{\rm{CO(7-6)}}^{'} = (3.3\pm 1.2)\times10^{9}\,$K$\,$\kms$\,$pc$^2$.
For reference, in a warmer scenario with $T_{\rm{gas}} \gtrsim 75\,$K the CMB correction becomes smaller ($f_{3\rightarrow2}\sim0.77$, $f_{7\rightarrow6}\sim0.85$), whereas for very cold gas approaching the CMB temperature the line contrast progressively vanishes.

\subsubsection{CO line ratios}\label{CO_line_ratios}

To infer a molecular gas mass from our CO(3–2) detection using the common approach, we must adopt an excitation correction--expressed as the brightness–temperature ratio $r_{31} = L'_{\rm CO(3-2)}/L'_{\rm CO(1-0)}$--to relate it to the ground-state CO luminosity. While observations of SFGs typically find $r_{J1}<1$, indicative of subthermal excitation, the actual value of $r_{31}$ depends sensitively on local gas conditions and therefore introduces an additional source of uncertainty.

We adopt $r_{31} = 0.84\pm0.26$, the median value reported by \citet{Riechers_2020} for the VLASPECS survey, a sample of $z = 2-3$ main-sequence galaxies. 
By doing so, we recover $L_{\rm{CO(1-0)}}^{'} = (3.4\pm 1.3)\times10^{10}\,$K$\,$\kms$\,$pc$^2$.
The $r_{31}$ we adopt is also in agreement with studies on somewhat different populations such as DSFGs, sub-millimeter galaxies (SMGs) and AGNs, and with the only direct measurement available at $z>6$, which yields $r_{31}\simeq1$ for the extreme starburst HFLS3 \citep[][]{Riechers_2013}, although the uncertainties on measured fluxes are significant and the scatter among individual sources is large \citep[e.g,][]{Sharon_2016, Frias_Castillo_2023, Prajapati_2025}. We also note that, although some high-$z$ sources show more excited gas conditions compared to local samples, some of these studies have tried to look for a redshift evolution (up to $z\lesssim6$) of $r_{31}$ but found none.

Thanks to the detection of both the $J = 3\rightarrow 2$ and $J = 7\rightarrow 6$ CO lines, we can compute the CO intrinsic brightness temperature ratio between these two transitions $r_{73} = L_{\rm{CO(7-6)}}^{'}/L_{\rm{CO(3-2)}}^{'} = 0.11\pm0.05$.\footnote{Since the CMB correction $f_{J \rightarrow J-1}$ we adopt is similar for both transitions, the ratio between the observed line luminosities measured directly is $r_{73} = 0.13\pm0.06$, which is consistent with the fiducial value.}

With only two CO fluxes associated with quite large uncertainties, the shape of the CO SLED remains uncertain. However, the measured excitation ratio is more consistent with star-forming systems than with extreme starburst or AGN-dominated sources, where more extreme ionizing radiation can populate higher $J$ levels, resulting in flatter CO SLEDs and significantly larger excitation ratios for mid- to high-$J$ CO lines \citep[e.g.,][]{van_der_Werf_2010, Carilli_Walter_2013, Riechers_2020}. More insight into the CO SLED of REBELS-25 can be obtained with radiative transfer modelling (see Sec.~\ref{TUNER}), where we find that in the best-fit model the CO SLED peaks at $J\simeq5$, consistent with moderately excited molecular gas conditions. 

\subsubsection{CO-to-H$_2$ conversion factor}\label{a_co_adopted}

To convert the CO(1--0) line luminosity into a molecular gas mass, we adopt the standard relation
\begin{equation}\label{eq_Mgas}
M_{\rm mol} = \alpha_{\rm{CO}}\, L_{\rm{CO(1-0)}}^{'},
\end{equation}
where $\alpha_{\rm{CO}}$ (in units of \msun(K$\,$\kms$\,$pc$^2$)$^{-1}$, omitted hereafter for readability) is the CO-to-H$_2$ conversion factor \citep[for a review, see][]{Bolatto_2013}. This conversion factor encapsulates the dependence of CO luminosity on physical conditions such as gas temperature, density, metallicity, and the interstellar radiation field strength, as supported by both theoretical and numerical studies \citep[e.g.,][]{Narayanan_2011, Feldmann_2012}. 

In the local Universe, $\alpha_{\rm{CO}}$ can vary by over an order of magnitude across galaxy populations. A Milky Way–like value of $\alpha_{\rm{CO}}\simeq4.3$ (including helium) is commonly adopted for main-sequence galaxies, while compact starbursts and (ultra-)luminous infrared galaxies (ULIRGs) typically show lower values, $\alpha_{\rm{CO}}\simeq1$, reflecting their enhanced CO surface brightness per unit molecular gas mass \citep[e.g.,][]{Daddi_2010a, Papadopoulos_2012, Carilli_Walter_2013, Tacconi_2020}.
Similar ULIRG-like conversion factors have also been inferred for several high-redshift SMGs based on dynamical mass constraints (e.g., \citealt{Hodge_2012, Riechers_2014, Riechers_2021a, Calistro-Rivera_2018}, c.f. \citealt{Dunne_2022}). 
Interestingly, \citet{Boogaard_2026} using the TUNER model (see Sec.~\ref{TUNER}) on GN20--a typical SMG at $z=4.055$--find $\alpha_{\rm{CO}}\simeq3.5$, higher than the dynamical constraints from \citet{Hodge_2012}, which indicated $\alpha_{\rm{CO}}=1.1\pm0.6$. 

At $z \gtrsim4$, direct observational constraints on $\alpha_{\rm CO}$ remain scarce and are absent at the EoR. 
For REBELS-25, we can exploit the dynamical mass inferred from the high-S/N, high-resolution [C\,\textsc{ii}] kinematic analysis \citep[$M_{\rm dyn} = 1.2^{+1.0}_{-0.6}\times 10^{11}\,$\msun;][]{Rowland_2024} to get an indication of an upper limit on the value of $\alpha_{\rm CO}$. Assuming the current best estimate for the stellar mass of the galaxy, a negligible contribution from dark matter within the probed region, and that the CO and [C\,\textsc{ii}] emission trace the same spatial extent (or, at least, that the CO emission is not more extended than the [C\,\textsc{ii}]), the dynamical mass implies $\alpha_{\rm CO}\lesssim3.5$ when adopting the CMB-corrected CO(1--0) luminosity derived above. This value is consistent with those expected for moderately enriched, main-sequence SFGs. 
On the other hand, the empirical metallicity–$\alpha_{\rm CO}$ relation from \citet{Genzel_2015} predicts $\alpha_{\rm CO}\simeq4.7$ for the sub-solar metallicity of REBELS-25 \citep[$Z\simeq0.85\,Z_\odot$;][]{Rowland_2024}, highlighting the systematic differences among existing approaches.

In the following, we adopt a fiducial value of $\alpha_{\rm CO}=3$, while also reporting results for $\alpha_{\rm CO}=1$ in the discussion. This range encompasses a range of plausible physical conditions inferred for REBELS-25, given the available observations and dynamical constraints. Independent insights on $\alpha_{\rm CO}$ are obtained through radiative transfer modelling of the CO and dust emission (Sec.~\ref{TUNER}). Additionally, a discussion on the impact of CMB on measured and inferred quantities is presented in Sec.~\ref{alpha_co_vs_redshift}.

\subsubsection{Molecular gas content}\label{Mgas_content}

Using Eq.$\,$\ref{eq_Mgas} and the CMB-corrected $L_{\rm{CO(1-0)}}^{'}$ derived in Sec.$\,$\ref{CO_line_ratios}, we compute a molecular gas mass of $M_{\rm{mol}} = (1.0\pm 0.4)\times 10^{11}\,\msun$ for $\alpha_{\rm{CO}}=3$. Adopting instead $\alpha_{\rm{CO}}=1$ yields $M_{\rm{mol}} = (3.4\pm 0.1)\times 10^{10}\,\msun$.
Under these assumptions, the values we derive imply the presence of an exceptionally massive molecular gas reservoir in a galaxy observed only $\simeq700$ Myr after the Big Bang, underscoring the rapid mass assembly occurring in REBELS-25.

Assuming a stellar mass of $\log(M_\star/\msun) = 9.30^{+0.12}_{-0.14}$ (Stefanon et al., in prep.), the inferred molecular gas masses correspond to gas fractions of $f_{\rm gas}=0.98\pm0.01$ and $0.94^{+0.03}_{-0.02}$ for $\alpha_{\rm CO}=3$ and $1$, respectively. In this case, the choice of the CO-to-H$_2$ conversion factor therefore has a relatively minor impact, yielding extreme gas fractions of $f_{\rm gas}\simeq0.95$ for typical $\alpha_{\rm CO}$ values adopted in the literature. 
Such high gas fractions likely reflect both the dominance of cold gas in the baryonic mass budget and a possible underestimation of the stellar mass due to substantial dust obscuration in the central region of the galaxy (see Sec.~\ref{mass_assembly}).

Together with the $M_{\rm dust}$, our CO-based estimates of $M_{\rm mol}$ imply $\delta_{\rm GDR} = (6^{+19}_{-5}) \times 10^2$ and $(2.2^{+6.5}_{-1.5}) \times 10^2$ for $\alpha_{\rm{CO}}=3$ and 1, respectively. These values are higher than those typically observed in the local Universe; however, they are similar to the $\delta_{\rm GDR}$ inferred for other galaxies at $z\gtrsim6.5$ \citep[][]{Algera_2026}, and--within the substantial uncertainties--consistent with locally calibrated scaling relations based on metallicity \citep[][]{Remy_Ruyer_2014, Galliano_2021}. 

Another relevant quantity we can derive is the gas depletion timescale $\tau_{\rm dep} = M_{\rm mol} / \mathrm{SFR}$, which measures how long the current gas reservoir could sustain star formation at the observed rate.
Using $\mathrm{SFR}_{\mathrm{UV+IR}} = 82^{+41}_{-21}\,\msun\, \mathrm{yr}^{-1}$ \citep[][]{Algera_2024b, Fisher_2026}, we obtain $\tau_{\rm dep} = 1.2^{+0.8}_{-0.6}\,$Gyr for $\alpha_{\rm CO}=3$, and $\tau_{\rm dep}=0.4^{+0.3}_{-0.2}\,$Gyr for $\alpha_{\rm CO}=1$.

\begin{figure}
 \includegraphics[width=\columnwidth,trim={0.48cm 0.6cm 0.1cm 0.05cm},clip]{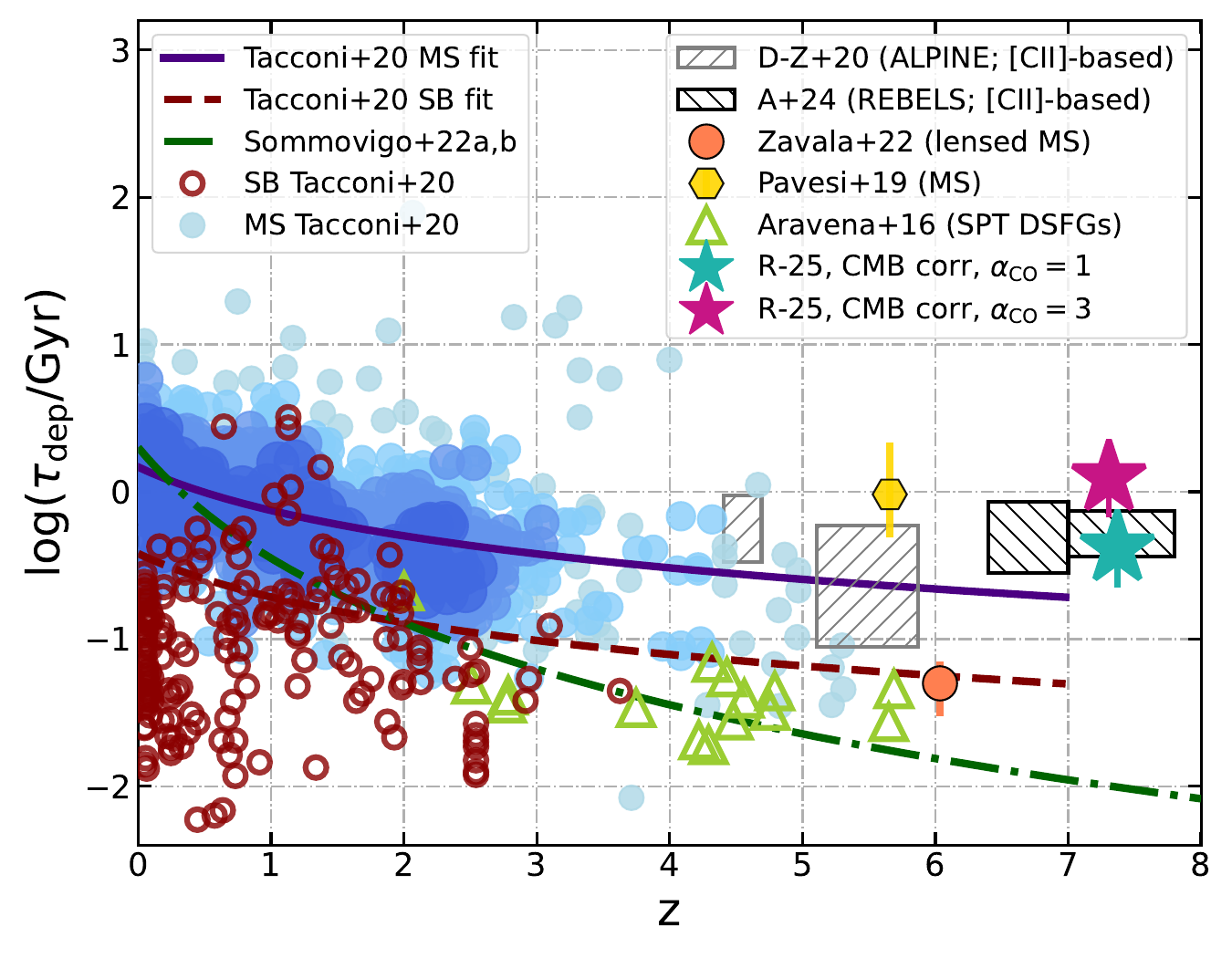}
 \caption{Molecular gas depletion timescale as a function of redshift for galaxies across cosmic time \citep[adapted from][]{Zavala_2022}. The large symbols mark the values derived for REBELS-25, assuming the CMB-corrected CO luminosity and $\alpha_{\rm CO}=3$ and 1 (magenta and aqua star, respectively); for clarity, the aqua star has been slightly offset in redshift to avoid overlapping error bars. Both values are broadly consistent with the extrapolation to $z\sim7$ of the best-fit relation for main-sequence (MS) galaxies from \citet{Tacconi_2020} (solid purple line). The dashed maroon line represents the corresponding fit from the same study for starburst (SB) galaxies. Blue and red circles indicate the individual sources in their respective MS and SB samples. The dash-dotted line shows the relation from \citet{Sommovigo_2022a, Sommovigo_2022b} based on the evolution of mean baryonic accretion rates onto dark matter halos in dark-matter–only simulations. The hatched gray boxes mark the typical range of values derived for MS galaxies at $z\sim4-6$ from [C\,\textsc{ii}]-based gas masses in the ALPINE survey \citep[][]{Dessauges-Zavadsky_2020}, while the black ones mark the range of values derived from [C\,\textsc{ii}]-based gas masses for the REBELS sample \citep[][]{Aravena_2024}. Additional high-redshift measurements from the literature are shown for comparison: the orange circle represents the lensed MS galaxy studied in \citet{Zavala_2022}, the yellow hexagon represents HZ10 from \citet{Pavesi_2019}, and green open triangles indicate SPT-selected DSFGs from \citet{Aravena_2016}.}
 \label{fig:depltime}
\end{figure}

The values we compute for REBELS-25 are comparable to the depletion timescale measured for HZ10 at $z=5.7$ \citep[$\tau_{\rm dep}=960^{+1200}_{-470}\,$Myr, derived adopting $\alpha_{\rm CO} = 4.5$][]{Pavesi_2019}, one of the few unlensed main-sequence galaxies at high redshift with a low-$J$ CO-based gas mass estimate. Our inferred depletion timescales are instead significantly longer than the value derived by \citet{Zavala_2022} for a lensed main-sequence galaxy at $z=6.03$ ($\tau_{\rm dep} = 50\pm10\,$Myr, adopting $\alpha_{\rm CO} = 1$). 
More broadly, the depletion timescales inferred for REBELS-25 are consistent with those inferred from [C\,\textsc{ii}]-based gas mass estimates (computed using the \citet{Zanella_2018} calibration) for the REBELS sample \citep[][]{Aravena_2024} and for SFGs at $z\sim4-6$ \citep[ALPINE survey;][]{Dessauges-Zavadsky_2020}. They are somewhat shorter than typical values for main-sequence galaxies at $z<2$, yet broadly agree with the extrapolation to $z\sim7$ of the empirical $\tau_{\rm dep}-z$ relation for main-sequence galaxies derived by \citet{Tacconi_2020} (see Fig.$\,$\ref{fig:depltime}).

In the same figure, we also show the $\tau_{\rm dep}$–$z$ relation from \citet{Sommovigo_2022a, Sommovigo_2022b}, which is based on the rapid redshift evolution of mean baryonic accretion rates onto dark matter halos in dark-matter–only simulations, reflecting the strong evolution of halo accretion timescales with redshift \citep[e.g.,][]{Fakhouri_2010, Correa_2015, Yung_2024}. By construction, this relation describes an average trend, and at any given epoch significant scatter and outliers are expected. While the predicted depletion times are somewhat shorter than those inferred for REBELS-25 at $z\sim7$, this difference can plausibly arise from both modelling assumptions (e.g. the lack of a distinction between different ISM phases) and from the specific physical conditions of REBELS-25. In particular, REBELS-25 is characterized by a relatively low dust temperature, which in this framework is associated with longer gas depletion timescales.

\subsection{Radiative transfer modelling}\label{TUNER}

\begin{figure*}
 \includegraphics[width=\textwidth,trim={0.1cm 0cm 0.1cm 0.1cm},clip]{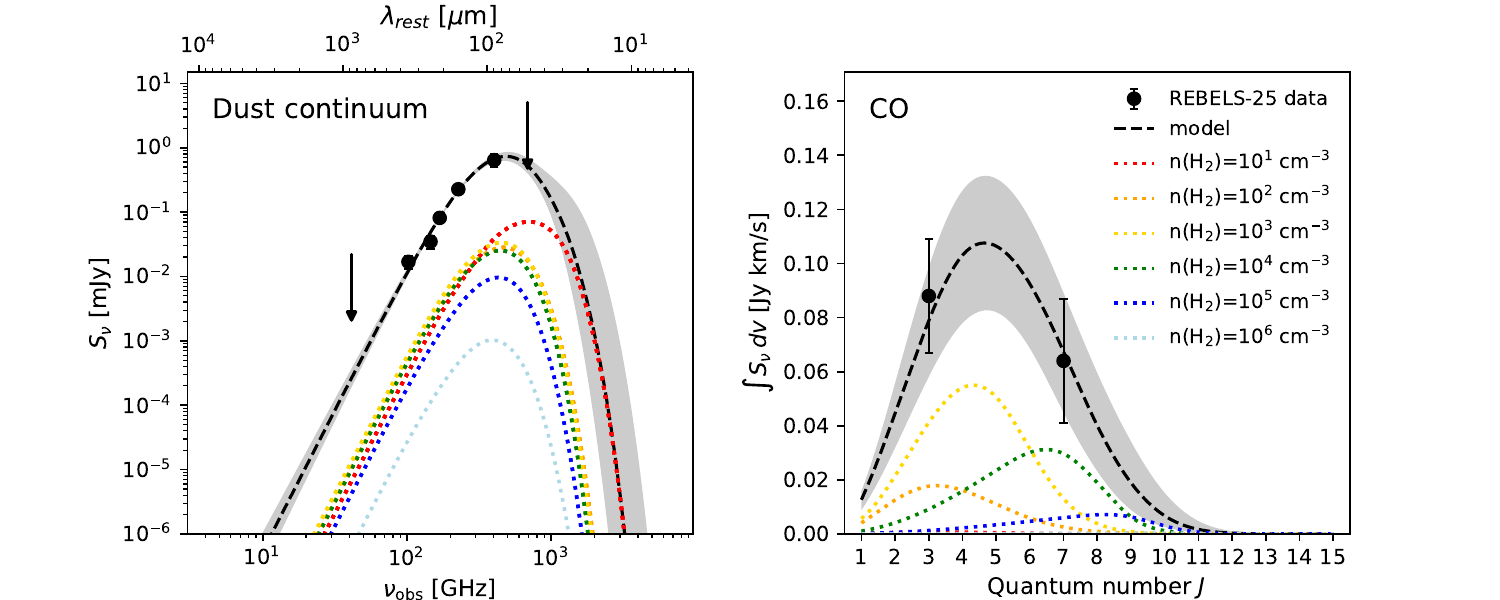}
 \caption{{Dust spectral energy distribution (SED; \textit{left panel}) and CO spectral line energy distribution (SLED; \textit{right panel}) of REBELS-25 derived by fitting the observed continuum and line fluxes (black points for detections, arrows for $1\sigma$ upper limits) self-consistently with the \texttt{TUNER} model. The black dashed lines represent the best-fit median model to the data given the priors, while the gray shaded regions show the 16th–84th percentile ranges. Coloured dotted lines represent the contribution of a few individual molecular gas density components, as described in the legend (and scaled up for visibility). The CO SLED in the model peaks around $J\simeq5$, consistent with moderately excited molecular gas conditions.}}
 \label{fig:tuner}
\end{figure*}

Using the Turbulent Non-Equilibrium Radiative Transfer model \citep[\texttt{TUNER};][and in prep.; see also \citealt{Weiss_2007, Harrington_2021}]{Boogaard_2026}, we jointly fit the CO and dust continuum measurements of REBELS-25 to infer the physical conditions of the molecular gas and obtain an estimate of $M_{\rm mol}$ independent of assumptions on $r_{31}$ and $\alpha_{\rm CO}$.

\texttt{TUNER} performs a self-consistent treatment of line and dust continuum emission based on a physically motivated description of the ISM. The model assumes that the molecular hydrogen gas density follows a lognormal distribution \citep[as would be in the case if turbulence sets the gas density distribution;][]{Padoan_2002, Krumholz_2005}, described by only two parameters: the mean density and the turbulent velocity width ($\Delta v_{\rm turb}$), which defines the width of the distribution. The CMB is naturally present in the equation of radiative transfer and also treated self-consistently.
It also assumes that the gas temperature has a power-law dependence on the gas density $T_{\rm kin} (n_{\rm H_2})\propto n_{\rm H_2}^{\gamma_T}$, where typically denser gas components are also colder, leading to ${\gamma_T}\leq 0$.
The excitation temperatures, optical depths and fluxes are then solved under the Large Velocity Gradient (LVG) approximation, with both the dust and line emission computed self-consistently, assuming a single [CO/H$_2$] abundance ratio and gas-to-dust mass ratio ($\delta_{\rm GDR}$) across the density distribution. 
The model is embedded in a Bayesian framework that enables posterior inference \citep[using \texttt{EMCEE};][]{Foreman-Mackey_2013} under flexible priors for all key parameters. 

We first fit the data, allowing the parameters to vary within physically motivated priors informed by observations of REBELS-25 and by values from the literature (see Appendix~\ref{sec:tuner_appendix}).
Since the CO-to-H$_2$ abundance ratio cannot be robustly constrained with the currently available data, we fix it to a value consistent with measurements of local star-forming galaxies \citep[e.g.,][]{Lacy_2017}, scaled to the sub-solar metallicity of REBELS-25 \citep[$Z\simeq 0.85\,Z_\odot$][]{Rowland_2026}, yielding $\log(\rm{[CO/H}_2])=-4.15$.
Figure~\ref{fig:tuner} shows the best-fit fiducial model to the dust continuum and CO measurements, while Fig.~\ref{fig:tuner_appendix} presents the corresponding corner plot of the posterior distributions. 

Several parameters are well constrained,\footnote{All uncertainties quoted from the \texttt{TUNER} model represent the 16th–84th percentiles of the posterior distribution.} including the dust temperature
($T_{\rm dust}=37^{+7}_{-4}\,$K)
and emissivity index
($\beta_{\rm dust}=2.39^{+0.21}_{-0.23}$),
both consistent with the results of \citet{Algera_2024b}. This good agreement depends on the dense sampling of the dust continuum emission, which does not leave much free room for different modified black body models to fit the SED. 
In terms of gas conditions, the model favours a relatively low volume-weighted median gas density 
($\log(\overline{n_{\rm H_2}}_{\rm , vol}[{\rm cm^{-3}}])=2.3^{+0.6}_{-0.4}$),
and returns a kinetic temperature posterior distribution with a median of $\simeq100\,$K but a peak at significantly lower values ($\simeq50\,$K), closer to the measured dust temperature. 
On the other hand, other parameters, such as $\gamma_T, \Delta v_{_{\rm turb}}$ and $\delta_{\rm GDR}$, remain less well constrained.
The total molecular gas mass derived from this model is $M_{\rm mol}= (1.8^{+1.0}_{-0.9})\times10^{11}\,\msun$.

We then explored alternative model setups, including different assumptions for the gas temperature floor above the CMB (see Appendix~\ref{sec:tuner_appendix} for details). 
We also tested the results obtained when decoupling the prior on $T_{\rm kin}$ from the one on $T_{\rm dust}$ -- whereas in the fiducial model these quantities were allowed to vary within $0.5 \leq T_{\rm kin}/T_{\rm dust} \leq 10$. 
We further explored models in which the CO abundance relative to H$_2$ is treated as a free parameter, adopting a uniform prior spanning the range $[{\rm CO/H_2}]=10^{-6}-10^{-3}$.
In this case, however, the posterior distribution of $[{\rm CO/H_2}]$ remains largely unconstrained, leading to broader posteriors for several other parameters.
Lastly, we tested models with fixed $\gamma_T=-0.25$ and $\Delta v_{\rm turb}=35$ and $78\,$\kms, where the latter value is based on the cold gas velocity dispersion measured from the [C\,\textsc{ii}] kinematic modelling \citep[in this case $\Delta v_{\rm turb}= 2\sqrt{2 \ln{2}}\, \overline{\sigma}_{\rm v}$, where $\overline{\sigma}_{\rm v}=33^{+9}_{-7}\,$\kms;][]{Rowland_2024}.
All runs yield consistent results for the total molecular gas mass, with median values in the range $M_{\rm mol}\simeq(1 - 8)\times10^{11}\,\msun$.

\section{Discussion}\label{sec:discussion}

With all methods explored in this work, we consistently infer a large molecular gas reservoir in REBELS-25. In particular, the molecular gas masses derived from the \texttt{TUNER} modelling are higher by a factor of $\simeq2-5$ compared to the estimates presented in Sec.~\ref{Mgas_content}, which are based on the standard approach commonly adopted at lower redshift. Unsurprisingly, the uncertainties associated with these gas mass estimates are substantial, especially for the radiative transfer modelling.
In contrast to Sec.~\ref{Mgas_content}, where fixed values of $r_{31}$ and $\alpha_{\rm CO}$ are adopted, \texttt{TUNER} infers both quantities self-consistently, without assuming LTE. Relaxing these assumptions naturally leads to broader posterior distributions, reflecting both the modest signal-to-noise ratio of the CO detections and the fact that the CO SLED is currently constrained by only two detected transitions. Although these measurements represent an unprecedented level of information for an unlensed star-forming galaxy at $z\gtrsim6$, the limited sampling of the SLED and the uncertainties on the measured CO fluxes inevitably leave significant degeneracies among the models that can reproduce the current data. Consequently, some of the gas-related properties and ultimately the molecular gas masses derived using \texttt{TUNER} carry substantial uncertainties. In this sense, the smaller uncertainties quoted in Sec.~\ref{Mgas_content} partly reflect the adopted assumptions rather than empirical constraints.

The higher median molecular gas mass returned by \texttt{TUNER} can partially be driven by a lower excitation correction than assumed in Sec.~\ref{CO_line_ratios}. From the model, we infer a CO excitation of $\simeq0.52$, which is lower, although broadly consistent with the value from the VLASPECS survey.

Another factor that can contribute to the larger gas masses derived using \texttt{TUNER} is a larger inferred CO-to-H$_2$ conversion factor. For the fiducial \texttt{TUNER} run, the conversion factor is $\alpha_{\rm CO}=9^{+6}_{-5}$,
which, however, cannot be directly compared to the common values adopted when using the standard approach (see Sec.~\ref{a_co_adopted}), as it is computed as the ratio between the gas mass and the observed luminosity against the background. 
From our best-fit model, we infer a CMB-correction factor for the CO(3--2) transition of $\sim0.6$. This is slightly larger than the value we adopted in Sec.~\ref{CMB_effect}, which was computed under the assumption of LTE and $T_{\rm gas}\simeq 40\,$K \citep[][]{da_Cunha_2013}.
Applying backwards this CMB correction to the \texttt{TUNER}-derived value yields an effective $\alpha_{\rm CO}\simeq5$, only slightly larger than the value measured for the Milky Way and similar to values measured for other star-forming galaxies. This result supports the use of an $\alpha_{\rm CO}>1$ for REBELS-25, while disfavouring a ULIRG-like value of $\alpha_{\rm CO}\sim1$ for this system.

The molecular gas masses inferred from the \texttt{TUNER} models formally exceed the dynamical constraints derived from the [C\,\textsc{ii}] kinematic and morphological modelling presented in \citet{Rowland_2024}. One way to reconcile these values is by considering a lower inclination for the galaxy when computing $M_{\rm dyn}$. REBELS-25 is observed close to face-on, and the \texttt{CANNUBI} fit yields a best-fit inclination of $(25\pm6)^\circ$. However, inclinations at such low angles are notoriously difficult to constrain in morphological modelling \citep[see e.g.,][]{Mancera_Pina_2022}, and the posterior distribution retains non-negligible probability toward even lower inclinations. If the system were instead inclined at $i\simeq19^\circ$ (i.e., within $1\sigma$ of the best-fit value), the dynamical mass would increase by a factor of $\simeq2$ to $M_{\rm dyn}\simeq2\times10^{11}\,$\msun, easing the tension.

\subsection{$\alpha_{\rm CO}$ in the early Universe}\label{alpha_co_vs_redshift}

Estimating the molecular gas mass from CO emission requires adopting an $\alpha_{\rm CO}$, which is known to be sensitively dependent on the physical conditions of the ISM. The values adopted in this work, $\alpha_{\rm CO}=1$ and $3$, should be regarded as galaxy-integrated, luminosity-weighted averages. In reality, $\alpha_{\rm CO}$ can vary substantially within a single system due to local variations in metallicity, gas density, turbulence, dust shielding, and the strength of the UV radiation field \citep[][]{Bolatto_2013, Narayanan_2012}.
At present, spatially resolved low-$J$ CO measurements for unlensed $z>6$ galaxies are beyond reach, and such integrated values represent the only feasible approach.

In the local Universe, $\alpha_{\rm CO}$ spans nearly an order of magnitude across environments, from $\alpha_{\rm CO}\simeq1$ in compact merger-driven ULIRGs, to $\simeq4.3$ in Milky Way–like disks, and even higher values in low-metallicity dwarf galaxies \citep[][]{Maloney_1988, Israel_1997}. Determining the appropriate conversion factor for high-redshift galaxies is even more challenging, as early galaxies are expected to host ISM conditions that differ markedly from those in the local Universe. Lower metallicities--more common in the early Universe, even for massive systems--imply lower carbon and oxygen abundances and, crucially, decreased dust shielding. This enhances photodissociation of CO, and consequently leads to an increase in $\alpha_{\rm CO}$ \citep{Wolfire_2010, Glover_2011, Narayanan_2012, Madden_2020}.
However, the dense, turbulent, and warm molecular gas often associated with high-$z$ starbursts pushes in the opposite direction, lowering $\alpha_{\rm CO}$ by enhancing CO excitation and increasing global CO optical depth.  
For REBELS-25, the very high neutral gas density inferred from the [O\,\textsc{i}]${145\,\mu{\rm m}}$ observations \citep[][]{Fudamoto_2025} qualitatively favours a relatively low $\alpha_{\rm CO}$, if similar conditions apply to the molecular gas.
The competition between these effects produces a complex, non-linear dependence of $\alpha_{\rm CO}$ on the underlying ISM properties.

Theoretical insight into this problem has been provided by simulations of galaxies in the Epoch of Reionization. \citet{Vallini_2018}, post-processing cosmological zoom-in simulations \citep[][]{Pallottini_2017a}, studied CO emission in a $z\sim6$ galaxy (\textit{Alth{\ae}a}) and found a mean conversion factor of $\langle\alpha_{\rm CO}\rangle = 1.54\pm0.9$, with relatively modest spatial variations across the disk. In their analysis, most of the scatter in $\alpha_{\rm CO}$ arises from local density inhomogeneities rather than metallicity or radiation field variations. 
While this value is consistent with those commonly adopted in the literature for high-redshift galaxies, it pertains to a single simulated system with specific characteristics and ISM conditions, and therefore cannot be taken as representative of the high-redshift galaxy population as a whole. In particular, \textit{Alth{\ae}a} has some characteristics that differ from REBELS-25: while having a comparable SFR ($\sim100\,\msun,\mathrm{yr}^{-1}$), it exhibits a lower [C\,\textsc{ii}] luminosity ($L_{\rm [CII]}\sim2\times10^{8}\,\lsun$) and metallicity ($Z\sim0.5\,\rm Z_\odot$), and a somewhat larger stellar mass ($\sim10^{10}\,\msun$),\footnote{Although we caution that the stellar mass of REBELS-25 might be underestimated due to heavy dust obscuration (see Sec.~\ref{mass_assembly}).} placing it on the SFMS at $z\sim6$. We also note that typically these simulated galaxies are more compact than some massive galaxies in the EoR, such as REBELS-25.
Despite these global differences and the substantial uncertainties associated with our measurements, the physical conditions of the molecular gas in the simulated systems in \texttt{SERRA} appear broadly similar to those inferred for REBELS-25. In particular, the typical molecular gas densities ($\log{(n_{\rm H_2} [{\rm cm}^{-3}])}\sim2.5$), gas turbulence ($\sim10\,$\kms), and gas kinetic temperature ($\sim100\,$K) \citep[][]{Pallottini_2017a, Pallottini_2019, Vallini_2018} are remarkably close to the values favoured by our modelling.

At high redshift, the impact of the CMB must also be carefully accounted for when interpreting CO observations \citep[][]{da_Cunha_2013, Zhang_2016}. The CMB affects CO emission in two ways. First, it acts as a background against which the line must be detected: for low-$J$ transitions whose excitation temperatures are comparable to $T_{\rm CMB}$, the line contrast is significantly reduced, suppressing the observed luminosity and, consequently, biasing gas mass estimates derived from uncorrected fluxes.  
On the other hand, as $T_{\rm CMB}$ increases with redshift, the gas and dust in galaxies are exposed to an increasingly warm radiation bath. This can excite the gas and alter the population of the CO rotational levels and therefore modify the shape of the CO SLED, especially for the lower levels.
These two effects act in opposite directions on the effective $\alpha_{\rm CO}$: the increased background leads to higher apparent $\alpha_{\rm CO}$ (due to lower observed CO fluxes), while the enhanced excitation decreases it (because certain CO transitions become easier to excite).
The net impact depends strongly on gas density and temperature, with cold, extended, low-density gas regions being dominated by the background effect, whereas for warmer, denser gas (as expected in compact starburst regions), the two effects can partially compensate.

An important implication is that, at sufficiently high redshift, the CO(1--0) transition may no longer provide the optimal probe of the bulk molecular gas reservoir. As the increasing CMB temperature raises the effective excitation floor, the CO level populations shift toward higher rotational states, such that the most populated level may occur at $J>1$. In this regime, CO(2--1) or CO(3--2) may be more favourably excited and easier to detect against the elevated background, while the commonly adopted scaling relations calibrated on CO(1--0) become increasingly uncertain and less applicable.

Using \texttt{TUNER}, we examine the CO partition function in our fiducial model for REBELS-25. We find that the $J=2$ level is marginally the most populated, followed closely by $J=1$ and $J=3$, which account for approximately 
$26\%$, $22\%$, and $21\%$
of the total CO population, respectively, with comparable contributions within the modelling uncertainties.
These considerations support the idea that the ground-state transition of CO may not always be the optimal probe for the molecular gas reservoir in galaxies during the EoR and beyond.

\subsection{[C\,\textsc{ii}] as a molecular gas mass tracer}\label{CII_as_Mmol_tracer}

\begin{figure}
 \includegraphics[width=\columnwidth,trim={0.4cm 0.6cm 0.9cm 0.3cm},clip]{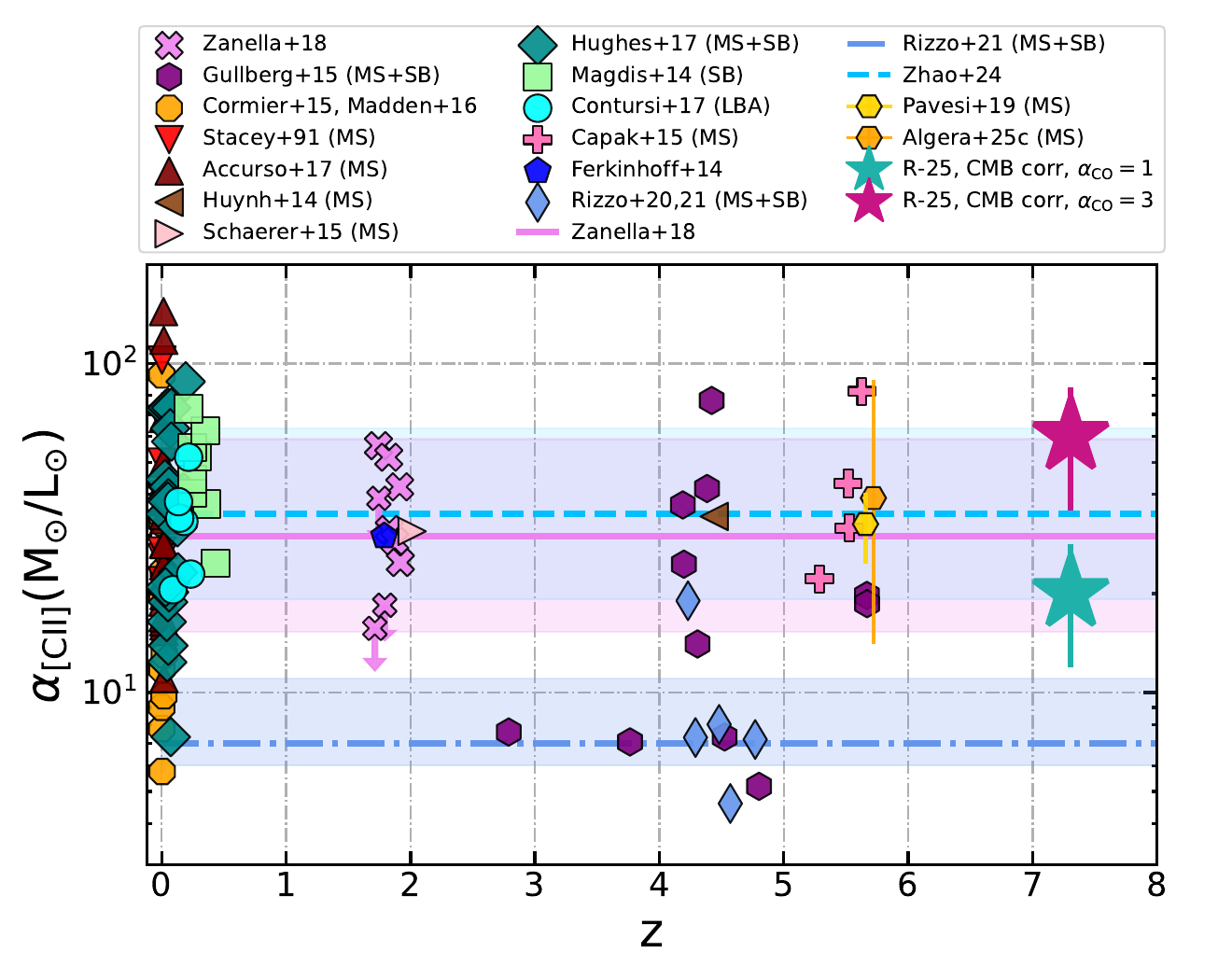}
 \caption{Evolution of the [C\,\textsc{ii}]-to-H$_2$ conversion factor ($\alpha_{\rm [C\,\textsc{ii}]}$) as a function of redshift. We include measurements drawn from the literature for main-sequence galaxies (MS), starburst galaxies (SB) and Lymann break analogues (LBA) \citep[][]{Stacey_1991, Ferkinhoff_2014, Huynh_2014, Magdis_2014, Capak_2015, Cormier_2015, Gullberg_2015, Schaerer_2015, Madden_2016, Accurso_2017, Contursi_2017, Hughes_2017, Zanella_2018, Rizzo_2020, Rizzo_2021}. The horizontal lines and shaded regions represent the mean and 1$\sigma$ scatter of the empirical $\alpha_{\rm [C\,\textsc{ii}]}$ calibrations from \citet{Zanella_2018} (pink, solid line) and \citet{Zhao_2024} (cyan, dashed line), and the median of the sample studied in \citet{Rizzo_2021} (cornflower blue, dash-dotted line). The large star symbols indicate the $\alpha_{\rm [C\,\textsc{ii}]}$ values we derive for REBELS-25 using the CMB-corrected CO luminosity and assuming either $\alpha_{\rm CO}=3$ (magenta) or $\alpha_{\rm CO}=1$ (aqua). The yellow and orange hexagons both refer to HZ10 \citep[see text;][]{Pavesi_2019, AHC_2025}, the latter has been slightly offset in redshift to avoid overlapping error bars. The values derived for REBELS-25 are broadly consistent with most low- and intermediate-redshift studies, suggesting that the [C\,\textsc{ii}] line remains an effective, though scattered, tracer of the molecular gas reservoir even at $z>7$.}
 \label{fig:alphacii}
\end{figure}

The combination of a low-$J$ CO detection and high–S/N spatially resolved [C\,\textsc{ii}] data for REBELS-25 provides a rare opportunity to test empirical and theoretical [C\,\textsc{ii}]–$M_{\rm mol}$ calibrations in the EoR. Using the molecular gas mass inferred from the CO measurements (Sec.~\ref{molecular_gas}) and the total [C\,\textsc{ii}] luminosity measured for the galaxy \citep[$L_{\rm [C\,\textsc{ii}]} = (1.7\pm0.2)\times10^9\,\msun$;][]{Hygate_2023}, we estimate an empirical [C\,\textsc{ii}]-to-H$_2$ conversion factor as $\alpha_{\rm [C\,\textsc{ii}]} = M_{\rm mol}/L_{\rm [C\,\textsc{ii}]}$.
This yields $\alpha_{\rm [C\,\textsc{ii}]} = (60 \pm 25)\,\msun / \lsun$ and $\alpha_{\rm [C\,\textsc{ii}]} = (20 \pm 8)\,\msun / \lsun$ for $\alpha_{\rm CO}=3$ and $1$, respectively. 

In Fig.~\ref{fig:alphacii}, we place the $\alpha_{\rm [C\,\textsc{ii}]}$ values derived for REBELS-25 in the context of measurements and literature compilations spanning the redshift range $0\lesssim z\lesssim6$. These studies encompass a variety of galaxy populations and employ conversion factors either anchored to CO-based gas mass estimates or derived from kinematic constraints \citep[][]{Stacey_1991, Ferkinhoff_2014, Huynh_2014, Magdis_2014, Capak_2015, Cormier_2015, Gullberg_2015, Schaerer_2015, Madden_2016, Accurso_2017, Contursi_2017, Hughes_2017, Zanella_2018, Pavesi_2019, Rizzo_2020, Rizzo_2021, Zhao_2024, AHC_2025}.
Within the uncertainties, our inferred $\alpha_{\rm [C\,\textsc{ii}]}$ values are consistent with many studies at low and intermediate redshift, as well as with the commonly adopted value of $31\,\msun/\lsun$ from \citet{Zanella_2018}. Adopting this reference value would imply a molecular gas mass of $M_{\rm mol}=(5.2^{+5.3}_{-2.6})\times10^{10}\,\msun$, which lies between our CO-based estimates obtained using $\alpha_{\rm CO}=3$ and 1.

We also compare our results to HZ10 at $z=5.7$. Using the [C\,\textsc{ii}] luminosity and the molecular gas mass inferred from the CO(2--1) detection reported by \citep[][who adopted $\alpha_{\rm CO}=4.5$]{Pavesi_2019}, and applying the same methodology adopted for REBELS-25, we derive a value of $\alpha_{\rm [C\,\textsc{ii}]} = (33\pm8)\,\msun/\lsun$ for this source. Using an alternative approach based on the dynamical mass and the total ISM mass budget (explicitly including both H\,\textsc{i} and H$_2$), \citet{AHC_2025} infer a [C\,\textsc{ii}]–to–total ISM mass conversion factor of $\alpha^{\rm ISM}_{\rm [C\,\textsc{ii}]} = (39^{+50}_{-25})\,\msun/\lsun$. Both estimates are comparable to our values derived from the CO(3--2) detection in REBELS-25.

These results suggest that [C\,\textsc{ii}] remains a viable, though scattered, alternative tracer of the molecular gas reservoir in the EoR. The scatter may partially reflect differences in methodologies used by different authors when analysing the data, but it also likely reflects an intrinsic dispersion driven by variations in ISM conditions and metallicity across galaxies.

\begin{figure*}
 \includegraphics[width=0.98\textwidth,trim={0.1cm 0.2cm 0.1cm 0.1cm},clip]{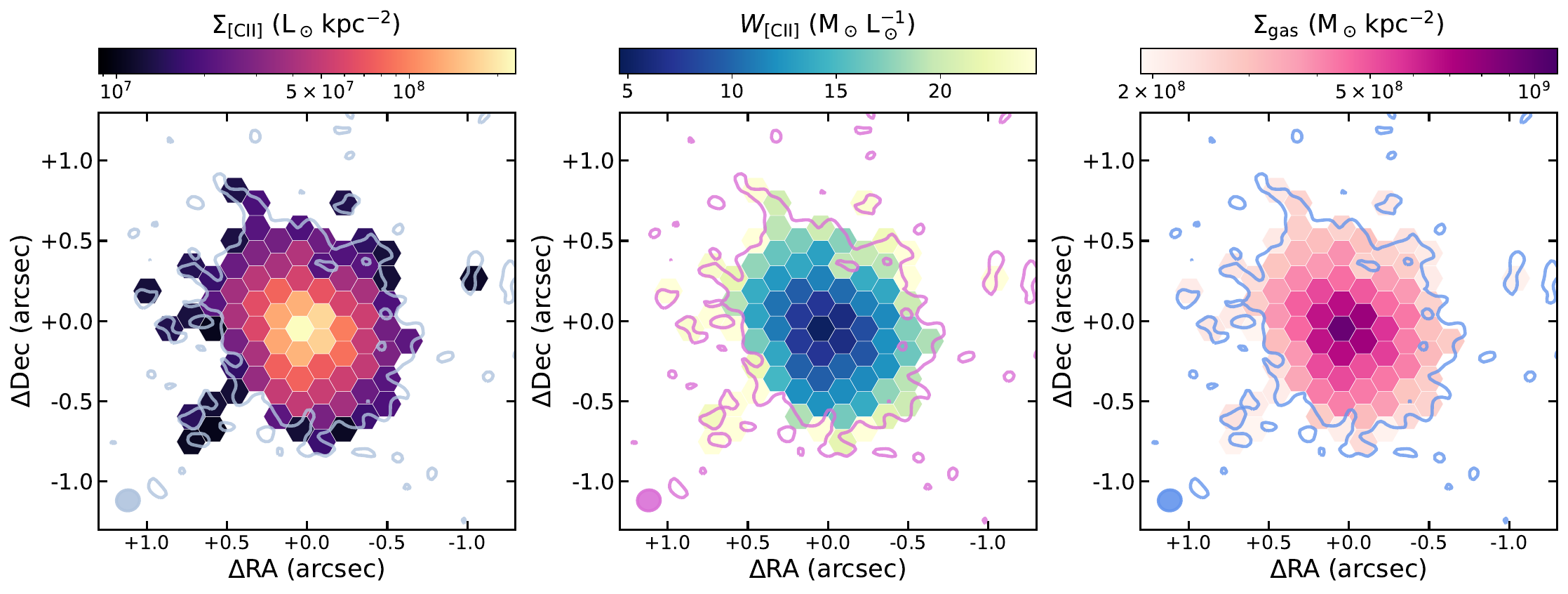}
 \caption{Resolved analysis of the [C\,\textsc{ii}]–$M_{\rm gas}$ relation in REBELS-25 following the prescription of \citet{Vallini_2025} based on simulated galaxies (see text for details). Maps of: [C\,\textsc{ii}] surface brightness ($\Sigma_{\rm [C\,\textsc{ii}]}$, in L$_{\odot}\,$kpc$^{-2}$) derived from the high–S/N, high–resolution ($\sim0.14\arcsec/700\,$pc) ALMA data (\textit{left panel}); resolved [C\,\textsc{ii}]-to-H$_2$ conversion factor ($W_{\rm [C\,\textsc{ii}]}$, in (M$_{\odot}\,$kpc$^{-2})/($L$_{\odot}\,$kpc$^{-2}$)) computed using Eq.~11 in V25 (\textit{central panel}); resulting gas surface density ($\Sigma_{\rm gas}$, in M$_{\odot}\,$kpc$^{-2}$; \textit{right panel}). In each panel, we show the synthesized beam of the [C\,\textsc{ii}] data as an ellipse in the bottom left corner and the contours of the [C\,\textsc{ii}] emission at a $2\sigma$ level.}
 \label{fig:Wcii}
\end{figure*}

In addition, we compare our findings to the work of \citealt{Vallini_2025} (hereafter V25), who used a set of $\sim100$ simulated galaxies from the \texttt{SERRA} cosmological zoom-in simulation suite \citep[][]{Pallottini_2022} to investigate [C\,\textsc{ii}] emission in simulated galaxies in the EoR. 
Their original integrated $\alpha_{\rm [C\,\textsc{ii}]}$–metallicity relation (Eq.~1 in V25) predicts relatively low conversion factors for systems with slightly sub-solar metallicity, as it is the case of REBELS-25, for which this relation would yield $\alpha_{\rm [C\,\textsc{ii}]} = 5.0^{+2.3}_{-1.6}\,\msun/\lsun$.
Part of this discrepancy likely arises from differences in the way metallicity is defined and measured in simulations versus observations. Observational metallicity estimates may be biased toward metal-rich, high–surface-brightness regions, particularly in heavily dust-obscured systems, as seems to be the case for the central region of REBELS-25 (Rowland et al., in prep.). In contrast, the metallicity adopted to derive Eq.~1 in V25 was averaged over a $25\,\mathrm{kpc}^2$ region around each simulated \texttt{SERRA} galaxy. Such a metallicity is not directly comparable to the metallicity inferred from observations of REBELS-25 and--also because \texttt{SERRA} galaxies are generally more compact than observed massive galaxies at similar redshift--likely includes a larger contribution from more pristine CGM gas, systematically shifting the average metallicity toward lower values \citep[see also][]{AHC_2025}.
When instead considering the metallicity measured within the [C\,\textsc{ii}] half-light radius of the simulated galaxies, a revised relation is obtained for the \texttt{SERRA} sample: 
\begin{equation}\label{Kohandel_prep}
\log(\alpha_{\rm [C\,\textsc{ii}]}) = -0.38 \log(Z/\text{Z}_\odot) + 1.12
\end{equation}
(Kohandel et al., in prep.). Applying this relation to REBELS-25 yields a predicted $\alpha_{\rm [C\,\textsc{ii}]} = 14^{+8}_{-5}\,\msun/\lsun$, closer to our empirical estimates.

We also apply the scaling relations that link $\alpha_{\rm [C\,\textsc{ii}]}$ to the effective [C\,\textsc{ii}] radius ($r_{\rm e,[C\,\textsc{ii}]}$; Eq.~12 in V25) and to the effective [C\,\textsc{ii}] surface brightness ($\Sigma_{\rm e,[C\,\textsc{ii}]} = L_{\rm [C\,\textsc{ii}]} / (2\pi r_{\rm e,[C\,\textsc{ii}]}^2)$; Eq.~13 in V25). Directly using $r_{\rm e,[C\,\textsc{ii}]}=2.2\pm0.3\,$kpc \citep[measured from the modelling of the high-resolution $\rm{[C\,\textsc{ii}]}$ data;][]{Rowland_2024}, we compute $\alpha_{\rm [C\,\textsc{ii}]} = 25^{+17}_{-10} \,\msun / \lsun$, consistent with both our direct estimate and the \citet{Zanella_2018} calibration. However, we note that this relation requires an extrapolation, as the [C\,\textsc{ii}] emission of REBELS-25 is more extended than that of the simulated galaxies in the study.
In contrast, the relation with $\Sigma_{\rm e,[C\,\textsc{ii}]}$ yields $\alpha_{\rm [C\,\textsc{ii}]} = 8^{+6}_{-3}\,$\msun/\lsun, a lower value that lies closer to the one predicted by the original integrated metallicity-based scaling.

Given the exceptionally high S/N and angular resolution of the available [C\,\textsc{ii}] data for REBELS-25, we also explore a spatially resolved estimate of the [C\,\textsc{ii}]-to-H$_2$ conversion factor ($W_{\rm [C\,\textsc{ii}]}$). Following the prescriptions of V25, we compute $W_{\rm [C\,\textsc{ii}]}$ maps using both the relation based solely on the [C\,\textsc{ii}] surface brightness ($\Sigma_{\rm [C\,\textsc{ii}]}$; Eq.~11 in V25), and the formulation that additionally includes a metallicity dependence (Eq.~10 in V25). In the metallicity-dependent case, due to the limited S/N of the JWST/IFU metallicity map \citep[][]{Rowland_2026}, we decided to adopt a single global metallicity for REBELS-25. Figure~\ref{fig:Wcii} illustrates the resulting $\Sigma_{\rm [C\,\textsc{ii}]}$, $W_{\rm [C\,\textsc{ii}]}$, and gas surface density ($\Sigma_{\rm gas}$) maps calculated using Eq.~11 in V25. Integrating the resolved gas surface density maps yields total cold gas masses of $M_{\rm gas}\approx 2\times10^{10}\,$\msun$\,$and $M_{\rm gas}\approx 9\times10^{9}\,$\msun, using Eq.~11 and 10, respectively.
These estimates are lower than the CO-based estimates, similarly to what we find using the relations relying on the integrated [C\,\textsc{ii}] emission.
We note that our approach currently relies on the integrated metallicity of REBELS-25. A spatially resolved metallicity map would enable a more robust test of these prescriptions for a resolved conversion factor. Achieving this, however, would require substantially deeper JWST observations, particularly in the heavily dust-obscured central regions.

Overall, REBELS-25 provides a crucial high-redshift benchmark for calibrating the [C\,\textsc{ii}]–$M_{\rm mol}$ relation. Our results highlight both the promise and the limitations of using [C\,\textsc{ii}] as a molecular gas tracer in the early Universe, emphasizing the importance of joint [C\,\textsc{ii}] and low-$J$ CO observations across a larger sample of galaxies to further investigate and test the use of empirical and theoretical conversion factors across a range of redshifts and ISM conditions.

\subsection{Comparison with alternative gas mass estimates from dust and [C\,\textsc{i}]}\label{Dust_and_CI}

We also estimate the cold gas mass using the long-wavelength dust-continuum calibration of \citet{Scoville_2016}. Under the optically–thin Rayleigh–Jeans (RJ) approximation, the gas mass can be written as 
$M_{\rm gas} = L_{\nu_{850\rm {\mu m}}} /  \alpha_{850\rm {\mu m}}$, where $L_{\nu_{850\rm {\mu m}}}$ is the rest-frame $850\,\mu$m monochromatic luminosity. For an observed flux density $S_{\nu_{\rm obs}}$ measured at the frequency corresponding to rest-frame $850\,\mu$m, i.e. $\nu_{\rm obs} = \nu_{850\rm {\mu m}}/(1+z)$, the luminosity is $L_{\nu_{850\rm {\mu m}}} = S_{\nu_{\rm obs}} 4\pi D_L ^2 / (1+z)$.
The coefficient $\alpha_{850\,\mu{\rm m}} = (6.7\pm1.7)\times10^{19}\,$erg$\,$s$^{-1}\,$Hz$^{-1}\,$M$_\odot^{-1}$ is the empirically calibrated luminosity-to-mass conversion derived from a combined sample of nearby spirals, local ULIRGs, and $z\sim2$ SMGs \citep{Scoville_2016}.

For REBELS-25, while the dust SED is relatively well constrained, there is no direct detection at rest-frame $850\,\mu$m. We therefore adopt the CMB-corrected observed-frame flux density corresponding to rest-frame $850\,\mu$m ($S_{\nu_{\rm obs} }= 0.76^{+0.41}_{-0.23}\, \mu$Jy), which is extrapolated from the best-MBB fit in \citet{Algera_2024b}. This yields a gas mass of $M_{\rm gas} = 9^{+5}_{-3} \times 10^{9}\,\msun$.
This estimate is a factor of $\approx 4-11$ lower than our CO-based gas mass measurements. However, it is important to note that the dust–based method hinges on several assumptions that are unlikely to hold for REBELS-25. For instance, the calibration of $\alpha_{850\mu{\rm m}}$ in \citet{Scoville_2016} assumes a quite high and fixed $\alpha_{\rm CO}$ value of 6.5 and implicitly a gas-to-dust ratio appropriate for more local and metal-rich galaxies. These assumptions are not necessarily valid for our massive, star-forming and heavily dust-obscured galaxy in the EoR. Variations in dust content, composition, and temperature and different gas conditions -- partly due to the elevated CMB temperature at $z=7.31$ -- can all affect the inferred gas mass. Thus, this comparison should be interpreted with caution, given the uncertain applicability of the underlying calibration to REBELS-25.

We can also place an upper limit on the molecular gas mass using our non-detection of the [C\,\textsc{i}](2–1) line. As discussed above for the CO transitions, at $z=7.31$ the CMB significantly affects the observed brightness of lines. We therefore first correct the upper limit derived from the observations for CMB attenuation. We assume $T_{\rm kin} = 40\,$K, consistent with the value adopted in our CO analysis (see Sec.~\ref{CMB_effect}). To estimate the magnitude of the suppression of the observed emission due to the CMB, we use the radiative-transfer calculations of \citet{Frias_Castillo_2025}. Their results explicitly show the attenuation of the ground-state [C\,\textsc{i}] line, which at the redshift of REBELS-25 is reduced by a factor of $\sim0.5$. This transition is expected to be more strongly affected by the CMB than the higher-excitation [C\,\textsc{i}] line considered in this work. Thus, adopting the same suppression factor provides a conservative correction. We therefore scale the observed luminosity upper limit as $L_{\rm [C\,\textsc{i}](2-1)}^{'} = L_{\rm [C\,\textsc{i}](2-1), \rm obs}^{'} / 0.5 < 3.2 \times10^{9}\,$K$\,$\kms$\,$pc$^2$. 
Following \citet{Weiss_2003}, the atomic carbon mass is then estimated as
\begin{equation}
\begin{split}
M_{\rm [C\,\textsc{i}]} &= 4.566 \times 10^{-4}\, Q(T_{\rm ex}) \, \frac{1}{5} \, e^{T_2/T_{\rm ex}} \, L'_{\rm [C\,\textsc{i}](2-1)} \; [{\rm M_{\odot}}],
\end{split}
\end{equation}
where $Q(T_{\rm ex}) = 1 + 3 e^{-T_1/T_{\rm ex}} + 5 e^{-T_2/T_{\rm ex}}$ is the [C\,\textsc{i}] partition function, and $T_1 = 23.6\,$K and $T_2 = 62.5\,$K are the energies of the excited states above the ground level. We adopt an excitation temperature of $T_{\rm ex} = 40\,$K, matching the assumed gas kinetic temperature, which is also consistent with the dust temperature measured in \citet{Algera_2024b} and in this work (Sec.~\ref{TUNER}). We note, however, that the [C\,\textsc{i}] emission can be sub-thermally excited \citep[][]{Papadopoulos_2022}, such that the LTE assumption may not strictly hold and impact the inferred carbon mass.
With these assumptions, we obtain $M_{\rm [C\,\textsc{i}]} < 5.2\times10^{6}\,$\msun.
To convert the carbon mass into a molecular gas mass, one must assume an abundance ratio $X_{\rm [C\,\textsc{i}]} = M_{\rm [C\,\textsc{i}]}/M_{\rm H_2}$. Different studies \citep[e.g.,][]{Valentino_2018, Boogaard_2020, Frias_Castillo_2025} find typical values of $X_{\rm [C\,\textsc{i}]} \sim(1-5)\times10^{-5}$, depending on the nature of the source, and on ISM conditions such as metallicity, density, and UV field strength. Reported abundances also depend on assumptions such as $\alpha_{\rm CO}$, gas-to-dust ratio or lensing correction. 
Adopting a conservative value of $X_{\rm [C\,\textsc{i}]} = 1 \times 10^{-5}$, we obtain a $3\sigma$ upper limit on the molecular gas mass of $M_{\rm mol} \lesssim 5.2\times10^{11}\,$\msun. 
This is less constraining than other available information on the galaxy--for instance, the dynamical mass inferred from [C\,\textsc{ii}] kinematics--and it is consistent with the gas masses inferred from CO, [C\,\textsc{ii}], and dust continuum emission. 
Nonetheless, it is informative since it hinges on a set of assumptions that differ from those underlying the other methods, and it is in tension with more extreme gas reservoir estimates that would require very low CO excitation or very large $\alpha_{\rm CO}$.

Conversely, we can use the [C\,\textsc{i}] upper limit to place tentative constraints on the implied atomic carbon abundance by requiring consistency with the molecular gas masses inferred from the CO and [C\,\textsc{ii}] measurements. Matching the CMB-corrected CO-based gas masses derived in Sec.~\ref{Mgas_content} implies $X_{\rm [C\,\textsc{i}]} < (5 \pm 2)\times10^{-5}$ and $<(1.5\pm0.6)\times10^{-4}$ when adopting $\alpha_{\rm CO}=3$ and 1, respectively. Requiring consistency with the [C\,\textsc{ii}]-based gas mass derived using the \citet{Zanella_2018} calibration yields $X_{\rm [C\,\textsc{i}]} < (1.0^{+1.0}_{-0.5})\times10^{-4}$. These upper limits fall within the broad range of atomic carbon abundances reported in the literature for star-forming galaxies \citep[e.g., ][]{Ikeda_2002, Weiss_2005, Jiao_2021}.

\subsection{Rapid mass assembly}\label{mass_assembly}

The detection of a large ($\sim10^{11}\,$\msun) molecular gas reservoir in REBELS-25, only $\simeq700\,$Myr after the Big Bang, is significant in the context of early galaxy evolution. The direct confirmation of such a large mass of cold gas at $z>7$ implies that even by this early epoch, some galaxies had accreted vast amounts of baryons, either through inflows, mergers, or a combination of both.

Observationally, gas fractions are known to increase steeply with redshift. High gas fractions ($f_{\rm gas}\sim0.5-0.8$) have been reported in several galaxies at $1.5\lesssim z \lesssim 4$ \citep[e.g.,][]{Daddi_2010a, Dessauges-Zavadsky_2015, Cassata_2020}, and empirical scaling relations from CO and dust surveys at $0\lesssim z\lesssim4$ show that the gas fraction evolves approximately as $f_{\rm gas} \propto (1+z)^{3}$ for main-sequence galaxies up to $z\sim3$ \citep[][]{Tacconi_2020}. 
At higher redshift, however, the scarcity of direct molecular gas detections prevents a clear assessment of whether this trend continues unabated, or if it begins to saturate \citep[see e.g.,][]{Aravena_2024}. 
Existing samples at high redshift are also often biased toward FIR-luminous systems, making it difficult to robustly distinguish between different proposed scaling relations. Extrapolating these commonly adopted relations to $z\sim7$ predicts typical gas fractions of $f_{\rm gas}\sim0.5-0.8$, while our inferred value of $f_{\rm gas}\simeq0.95$ places REBELS-25 at the extreme gas-rich end of the currently known high-redshift population. Our inferred gas fraction appears to be in better agreement with the scaling relations proposed by \citet{Liu_2019}. In their work, the authors compiled a large galaxy sample by mining the ALMA archive in the COSMOS deep field and combining these sub-mm detected sources with previous CO- and dust-based molecular gas measurements spanning a broad range of redshifts and galaxy properties. Their analysis suggested higher gas fractions and longer depletion times in lower-mass systems at high redshift. However, some caution is warranted when interpreting this comparison, as REBELS-25 would fall in a ($z, M_\star, \rm {SFR}$) parameter space that is not directly sampled by their compilation.

In this context, the detection presented here provides an additional constraint on the evolution of gas fractions in the early Universe, particularly in a regime of relatively low stellar masses and high redshift where observational constraints remain extremely limited. At the same time, REBELS-25 still represents a single object, and significantly larger samples of direct molecular gas detections at $z>7$ will be required to robustly constrain the evolution of gas fractions and distinguish between different proposed scaling relations.

A potential tension with the high $f_{\rm gas}$ we derive emerges when considering the near-solar metallicity measured for REBELS-25 \citep[12+log(O/H)=$8.62\pm0.17$;][]{Rowland_2024}. 
In simple galaxy evolution scenarios, star formation simultaneously enriches the ISM and depletes the gas reservoir, implying that systems with high metallicities are expected to have already converted a significant fraction of their gas into stars. Even in a closed-box model--which neglects any inflows or outflows and maximizes metallicity at fixed gas fraction--our measurements appear difficult to reconcile, suggesting that either the gas mass or metallicity is overestimated, the stellar mass is significantly underestimated, or a combination of these effects (see also the discussion in \citealt{Algera_2026}).

Several considerations may alleviate this inconsistency. First, the stellar mass estimate of REBELS-25, derived from integrated SED fitting including JWST/NIRSpec data (Stefanon et al., in prep.), may be underestimated due to substantial dust obscuration in the central region of the source. Moreover, studies employing spatially resolved SED fitting--which partially mitigates outshining by young stellar populations--often recover systematically larger stellar masses \citep[e.g.,][]{Sorba_2018, Gimenez-Arteaga_2023, Gimenez-Arteaga_2024}, although recent works suggest that outshining effects become less severe above $\log(M_\star/\rm M_\odot)\gtrsim9$ \citep[][]{Li_2024, Lines_2025}. Deeper, high-resolution observations extending to redder wavelengths--particularly sampling the rest-frame $\sim1\,\mu$m emission with JWST/MIRI--will be crucial for refining the stellar mass measurement.
In support of this interpretation, the exceptionally high dust-to-stellar mass ratio of REBELS-25 \citep[$\log({\rm DTS})\simeq-1.1$;][]{Algera_2026} independently points toward an underestimated stellar mass. Such a high DTS is striking even in the context of other high-redshift galaxies and lies at the upper edge of the range expected from current models of dust production and growth \citep[][]{Ferrara_2025}. 
Notably, adopting the pre-JWST stellar mass estimates would instead yield a more moderate value of $\log({\rm{DTS}})\simeq-2.1$. Similar DTS values have been reported in lower-redshift dusty star-forming galaxies \citep[e.g.,][]{Donevski_2020}, although such comparisons should be treated with caution given the substantial differences in redshift and typical stellar masses between those systems and REBELS-25.

While an underestimation of the stellar mass likely plays a dominant role in driving the inferred DTS, we note that, despite the dust SED being well constrained, the dust mass could also be mildly overestimated. If part of the dust emission is optically thick, it could mimic a lower effective dust temperature and impact the inferred dust mass. Such effects may partially arise from dust properties that differ from those in local galaxies, for instance, through uncertainties in the dust opacity, which remains poorly constrained at high redshift.
Nevertheless, even adopting the higher stellar mass estimate previously proposed by \citet{Topping_2022} ($M_\star=1.9^{+0.5}_{-0.8}\times10^{10}\,\msun$), yields a substantial gas fraction of $f_{\rm gas}\simeq0.65-0.85$, depending on the adopted $\alpha_{\rm CO}$, reinforcing the conclusion that the galaxy remains gas dominated.

Uncertainties in the metallicity measurement may also play a significant role. Metallicity estimates based on rest-frame optical emission lines are luminosity-weighted and preferentially trace compact, actively star-forming regions, potentially biasing the inferred metallicity high relative to the galaxy-wide average. 
In addition, systematic uncertainties associated with the choice of diagnostics and calibrations are quite substantial. Owing to the modest S/N of the NIRSpec data and the lack of H$_\alpha$ coverage at $z>7$, \citet{Rowland_2026} derive the fiducial metallicity of REBELS-25 using the $R3$ = [O\,\textsc{iii}]$_{5007}$/H$_{\beta}$ ratio and the calibration of \citet{Sanders_2024}. As discussed in their section 4.2.1, adopting alternative diagnostics and calibrations yields a wide range of values, spanning $8.02\pm0.27 \leq$12+log(O/H)$\leq 8.71\pm0.20$ ($0.21\lesssim Z/\rm Z_\odot \lesssim 1.05$). 
This range could partially reconcile the high inferred gas fraction. 
We also note that a potential additional source of uncertainty is the presence of a weak AGN, which could bias emission-line ratios such as $R3$ and thus affect the inferred metallicity; however, current data do not allow us to robustly assess this possibility. Observations of auroral lines such as [O\,\textsc{iii}]$_{4363}$ would enable direct electron-temperature metallicity measurements \citep[e.g.,][]{Curti_2023}, providing a more robust benchmark.

Taken together, these uncertainties suggest a less extreme, yet still strongly gas-dominated scenario, in which a substantial fraction of the baryonic mass of REBELS-25 is in the form of cold gas. 
Even if somewhat overestimated, the $f_{\rm gas}$ of this galaxy is likely very high, potentially pointing to a short-lived phase of rapid mass assembly, during which cold accretion streams may deliver gas at rates comparable to, or exceeding, the SFR, sustaining high gas fractions and short duty cycles of intense star formation. In this scenario, REBELS-25 may represent a transient stage in which gas inflow temporarily outpaces star formation, producing a gas-dominated system with a potential for future substantial stellar mass growth. 
This gas-dominated scenario is also qualitatively consistent with its large degree of rotational support, as indicated by the high $V_{\rm{rot, max}}/\overline{\sigma}$ inferred from the [C\,\textsc{ii}] kinematic modelling, which points to a dynamically cold, rotation-dominated disk with high specific angular momentum. In the local Universe, galaxies with higher gas fractions are observed to preferentially reside in higher-angular-momentum systems \citep[e.g.,][]{Mancera_Pina_2021}, suggesting that the dynamical state of REBELS-25 is at least broadly compatible with its inferred gas-rich nature.

Interestingly, despite the quite high SFR of REBELS-25 (${\rm SFR_{UV+IR}} = 82^{+41}_{-21}\,{\rm M_\odot\,yr^{-1}}$), its inferred depletion time ($\tau_{\rm dep}\simeq0.4-1.2\,$Gyr) seems to be more consistent with the extrapolation of the redshift trends derived for main-sequence galaxies at $0\lesssim z \lesssim 4$ than in starbursts \citep[see Fig.~\ref{fig:depltime};][]{Tacconi_2020}. This contrasts, for instance, with the value derived from CO(2--1) for the $z=6.03$ main-sequence galaxy analysed by \citet{Zavala_2022}, where the depletion timescale measured for their source resembles that of extreme local starbursts (see Fig.~\ref{fig:depltime}). 
However, although their source lies on the SFMS, it appears to be more akin to some massive DSFGs, and therefore may not be directly comparable to REBELS-25 in terms of its dust and molecular gas properties. 

An additional insight into the evolution of REBELS-25 comes from its gas-to-dust ratio (see also \citealt{Algera_2026}). Our CO-based measurements, though carrying large uncertainties, yield values higher than those typical of local massive galaxies. The value obtained when adopting $\alpha_{\rm CO}=1$, $\delta_{\rm GDR} = (2.2^{+6.5}_{-1.5})\times10^2$, is broadly consistent with locally calibrated scaling relations (using the metallicity-based relation by \citet{Remy_Ruyer_2014} together with the metallicity of REBELS-25, we would expect $\delta_{\rm GDR} = (1.4^{+0.8}_{-0.5})\times10^2$). The values derived when assuming $\alpha_{\rm CO}=3$ or when using the \texttt{TUNER}-based $M_{\rm mol}$ are significantly larger, and more similar to what has been observed in several other $z>6$ galaxies.
However, high-redshift measurements of $\delta_{\rm GDR}$ exhibit substantial scatter. For example, our estimates for REBELS-25 are higher than the value found for the lensed $z=6.03$ galaxy studied by \citet{Zavala_2022}, who reported $\delta_{\rm GDR}=105\pm40$, yet they remain well below the extremely high ratio recently obtained for the main-sequence galaxy AD1689-zD1 at $z=7.13$ \citep[$\delta_{\rm GDR}=(2.0^{+1.2}_{-0.7})\times10^3$;][]{Heintz_2025}. This wide range likely reflects diversity in ISM enrichment histories and dust production pathways among galaxies in the EoR.

\begin{table*}
\centering
\caption{Summary of observational measurements and gas mass estimates for REBELS-25.
The table lists synthesized beamwidth ($\theta_{\rm HPBW}$), observed line luminosities, molecular and total cold gas mass estimates derived using different tracers and assumptions (all CO-based estimates include the CMB correction; see Sec.~\ref{CMB_effect}), and the resulting [C\,\textsc{ii}]–to–H$_2$ conversion factors.
}
\label{tab:results}
\begin{tabular}{llc}
\hline\hline
Quantity & Method / Assumptions & Value \\
\hline
\multicolumn{3}{l}{\textit{Observational properties}} \\
\hline
$\theta_{\rm HPBW,\,CO(3-2)}$ & VLA Q-band data (Sec.~\ref{subsec:VLAdata}) & $2.26\arcsec \times 1.86\arcsec$ \\
$\theta_{\rm HPBW,\,CO(7-6)}$ & ALMA Band 3 data (Sec.~\ref{subsec:ALMAdata}) & $1.00\arcsec \times 0.94\arcsec$ \\
$L'_{\rm CO(3-2),obs}$ & Gaussian fitting to the line profile (Sec.~\ref{line_luminosities}) & $(1.6\pm0.4)\times10^{10}$ K km s$^{-1}$ pc$^{2}$ \\
$L'_{\rm CO(7-6),obs}$ & Gaussian fitting to the line profile (Sec.~\ref{line_luminosities}) & $(2.1\pm0.8)\times10^{9}$ K km s$^{-1}$ pc$^{2}$ \\
$L'_{\rm [C\,\textsc{i}](2-1),obs}$ & $3\sigma$ upper limit (Sec.~\ref{line_luminosities}) & $<1.6\times10^{9}$ K km s$^{-1}$ pc$^{2}$ \\
\hline
\multicolumn{3}{l}{\textit{Molecular and total cold gas mass estimates}} \\
\hline
$M_{\rm mol}$ & CO(3--2), $r_{31}$, $\alpha_{\rm CO}=3$ (Sec.~\ref{Mgas_content}) & $(1.0\pm0.4)\times10^{11}\,\msun$ \\
$M_{\rm mol}$ & CO(3--2), $r_{31}$, $\alpha_{\rm CO}=1$ (Sec.~\ref{Mgas_content}) & $(3.4\pm0.1)\times10^{10}\,\msun$ \\
$M_{\rm mol}$ & Radiative transfer modelling (\texttt{TUNER}; Sec.~\ref{TUNER}) & $(1.8^{+1.0}_{-0.9})\times10^{11}\,\msun$ \\
$M_{\rm mol}$ & [C\,\textsc{ii}], $\alpha_{\rm [C\,\textsc{ii}]}=31\,\msun/\lsun$ (\citealt{Zanella_2018}; Sec.~\ref{CII_as_Mmol_tracer}) & $(5.2^{+5.3}_{-2.6})\times10^{10}\,\msun$ \\
$M_{\rm gas}$ & Resolved [C\,\textsc{ii}] (\citealt{Vallini_2025}; Sec.~\ref{CII_as_Mmol_tracer}) & $\approx2\times10^{10}\,\msun$ \\
$M_{\rm gas}$ & Dust continuum (\citealt{Scoville_2016}; Sec.~\ref{Dust_and_CI}) & $9^{+5}_{-3}\times10^{9}\,\msun$ \\
$M_{\rm mol}$ & [C\,\textsc{i}] upper limit, $X_{\rm [C\,\textsc{i}]}\sim10^{-5}$ (Sec.~\ref{Dust_and_CI}) & $\lesssim5.2\times10^{11}\,\msun$ \\
\hline
\multicolumn{3}{l}{[C\,\textsc{ii}]-to-H$_2$ \textit{conversion factors}} \\
\hline
$\alpha_{\rm [C\,\textsc{ii}]}$ & CO-based, $\alpha_{\rm CO}=3$ (Sec.~\ref{CII_as_Mmol_tracer}) & $(60\pm25)\,\msun/\lsun$ \\
$\alpha_{\rm [C\,\textsc{ii}]}$ & CO-based, $\alpha_{\rm CO}=1$ (Sec.~\ref{CII_as_Mmol_tracer}) & $(20\pm8)\,\msun/\lsun$ \\
\hline
\end{tabular}
\end{table*}

\section{Summary and conclusions}\label{sec:conclusions}

Utilizing VLA Q-band and ALMA Band 3 observations of REBELS-25, a star-forming galaxy at $z=7.31$, we report the detection of CO(3--2) and CO(7--6), along with an upper limit on [C\,\textsc{i}](2--1). This constitutes the first detection of low-$J$ CO emission in a star-forming galaxy at $z>7$. We correct the observed CO(3--2) flux for CMB-induced suppression following \citet{da_Cunha_2013} and adopt an excitation ratio $r_{31}=0.84\pm0.26$ \citep[][]{Riechers_2020}. Assuming $\alpha_{\rm CO}=3$ and $1$, respectively, we derive molecular gas masses of $M_{\rm{mol}} = (1.0\pm 0.4)\times 10^{11}\,\msun$ and $M_{\rm{mol}} = (3.4\pm 0.1)\times 10^{10}\,\msun$.

We self-consistently model CO and dust emission with the novel radiative transfer code \texttt{TUNER} \citep[][and in prep.; see also \citealt{Weiss_2007, Harrington_2021}]{Boogaard_2026}. The resulting molecular gas mass,
$M_{\rm mol}= (1.8^{+1.0}_{-0.9})\times10^{11}\,\msun$,
is agnostic to assumptions about $r_{31}$ and $\alpha_{\rm CO}$, and independently confirms the presence of a large molecular gas reservoir. The model fits for several galaxy properties, among which we find $T_{\rm dust}=37^{+7}_{-4}\,$K and $\beta_{\rm dust}=2.39^{+0.21}_{-0.23}$
(consistent with \citealt{Algera_2024b}), and a relatively low volume-weighted median gas density $\log(\overline{n_{\rm H_2}}_{\rm , vol}[{\rm cm^{-3}}])=2.3^{+0.6}_{-0.4}$.
The best-fit CO SLED peaks at $J\simeq5$, suggesting moderate excitation of the molecular gas, supported by the relatively weak observed CO(7--6) luminosity.
A summary of the main observational measurements and inferred physical properties is provided in Table~\ref{tab:results}.

Using the CO(3--2)-based gas masses, we infer [C\,\textsc{ii}]–to–gas conversion factors of $\alpha_{\rm [C\,\textsc{ii}]}=(60\pm25)\,{\rm M_\odot/L_\odot}$ and $(20\pm8)\,{\rm M_\odot/L_\odot}$ for $\alpha_{\rm CO}=3$ and $1$, respectively. Within the large scatter, these values are broadly consistent with studies at lower redshift and with the commonly adopted calibration by \citet{Zanella_2018}, suggesting [C\,\textsc{ii}] remains a viable alternative cold gas tracer in the EoR. 
We further compare our CO-based values to recent results from studies that post-processed and analysed cosmological zoom-in simulations \citep[][]{Vallini_2025}. We first test prescriptions based on integrated galaxy properties, and then exploit the exceptionally high-S/N and high-resolution [C\,\textsc{ii}] observations available for REBELS-25 to explore the use of a resolved [C\,\textsc{ii}]-to-H$_2$ conversion factor $W_{\rm [C\,\textsc{ii}]}$.
Depending on the specific method adopted, in some cases we derive values that are consistent with our findings, while in other cases simulations would suggest lower cold gas masses for REBELS-25.
Additionally, we derive a molecular gas mass from dust-continuum emission (following \citealt{Scoville_2016}) and place an upper limit using the [C\,\textsc{i}](2--1) non-detection, and compare those estimates with our direct CO-based measurements.

The direct confirmation of such a massive cold gas reservoir only $\simeq700\,$Myr after the Big Bang indicates that REBELS-25 had already assembled a substantial baryonic component. 
Combined with the properties of the galaxy, estimated from a rich set of ancillary data, our measurements imply an extreme gas fraction of $f_{\rm gas}\simeq 0.95$, which indicates that this system is gas-dominated, and a depletion timescale of $\tau_{\rm dep}\simeq0.4-1.2\,$Gyr, which we show to be broadly consistent with the extrapolation of empirical trends derived at low- and intermediate-redshift \citep[][]{Tacconi_2020}. Additionally, we derive a gas-to-dust ratio of $\delta_{\rm GDR} \simeq 200-600$.

With this study, we demonstrate the feasibility of detecting CO(3--2) emission in a bright star-forming galaxy deep within the EoR.vExpanding the sample of sources with low-$J$ CO and [C\,\textsc{ii}] detections at $z>7$ is essential for robustly constraining the cold gas content and testing alternative molecular gas tracers in the early Universe. 
Upcoming and next-generation facilities such as the SKA and ngVLA will be transformative, enabling the detection of low-$J$ CO in more typical galaxies at high redshift and dramatically increasing the number of systems for which molecular gas content can be directly measured. Such observations will provide crucial insights into star formation efficiency, ISM enrichment, and the mass assembly histories of the early galaxies.

\section*{Acknowledgements}


We thank the referee for their constructive comments and suggestions, which helped improve the paper.
KC acknowledges Piyush Sharda and Celine Greis for useful discussions during the early stages of this work.
KC acknowledges support from a grant from the Leiden University Fund / Fonds van Trigt. 
JAH and LAB acknowledge support from the ERC Consolidator Grant 101088676 (“VOYAJ”). 
LAB acknowledges support from the Dutch Research Council (NWO) under grant VI.Veni.242.055 (\url{https://doi.org/10.61686/LAJVP77714}). 
HSBA gratefully acknowledges support from Academia Sinica through grant AS-PD-1141-M01-2. 
DR gratefully acknowledges support from the Collaborative Research Center 1601 (SFB 1601 sub-projects C1, C2, C3, and C6) funded by the Deutsche Forschungsgemeinschaft (DFG) – 500700252. 
MA is supported by FONDECYT grant number 1252054, and gratefully acknowledges support from ANID Basal Project FB210003,  ANID MILENIO NCN2024\_112 and ANID + Vinculaci\'on Internacional + FOVI250261. 
IDL acknowledges support through funding from the European Research Council (ERC) under the European Union's Horizon 2020 research and innovation program DustOrigin (ERC-2019-StG-851622), and from the Flemish Fund for Scientific Research (FWO-Vlaanderen) through the research project G0A1523N.
PEMP is funded by the Dutch Research Council (NWO) through the Veni grant VI.Veni.222.364.
MR was supported by the NWO Veni project "\textit{Under the lens}" (VI.Veni.202.225).
LV acknowledges support from the INAF Minigrant "RISE: Resolving the ISM and Star formation in the Epoch of Reionization" (Ob. Fu. 1.05.24.07.01).
The National Radio Astronomy Observatory and Green Bank Observatory are facilities of the U.S. National Science Foundation operated under cooperative agreement by Associated Universities, Inc.
This research made use of \texttt{Astropy} \citep[][]{Astropy_2013, Astropy_2013_2}, \texttt{Matplotlib} \citep[][]{Hunter_2007_matplotlib} and \texttt{NumPy} \citep[][]{Harris_2020_numpy}.

\section*{Data Availability}


This paper makes use of the following ALMA data: ADS/JAO.ALMA$\#$2021.1.01495.S. The data are available in the ALMA archive https://almascience.eso.org/aq/ and can be accessed with the project code: 2021.1.01495.S. ALMA is a partnership of ESO (representing its member states), NSF (USA) and NINS (Japan), together with NRC (Canada), MOST and ASIAA (Taiwan), and KASI (Republic of Korea), in cooperation with the Republic of Chile. The Joint ALMA Observatory is operated by ESO, AUI/NRAO, and NAOJ. The VLA observations used in this article are available in the NRAO archive https://data.nrao.edu/portal/$\#$/ and can be accessed with the project code: 21A-335.



\bibliographystyle{mnras}
\bibliography{catalog} 




\appendix

\section{\texttt{TUNER} modelling}\label{sec:tuner_appendix}

\begin{figure*}
 \includegraphics[width=1\textwidth]{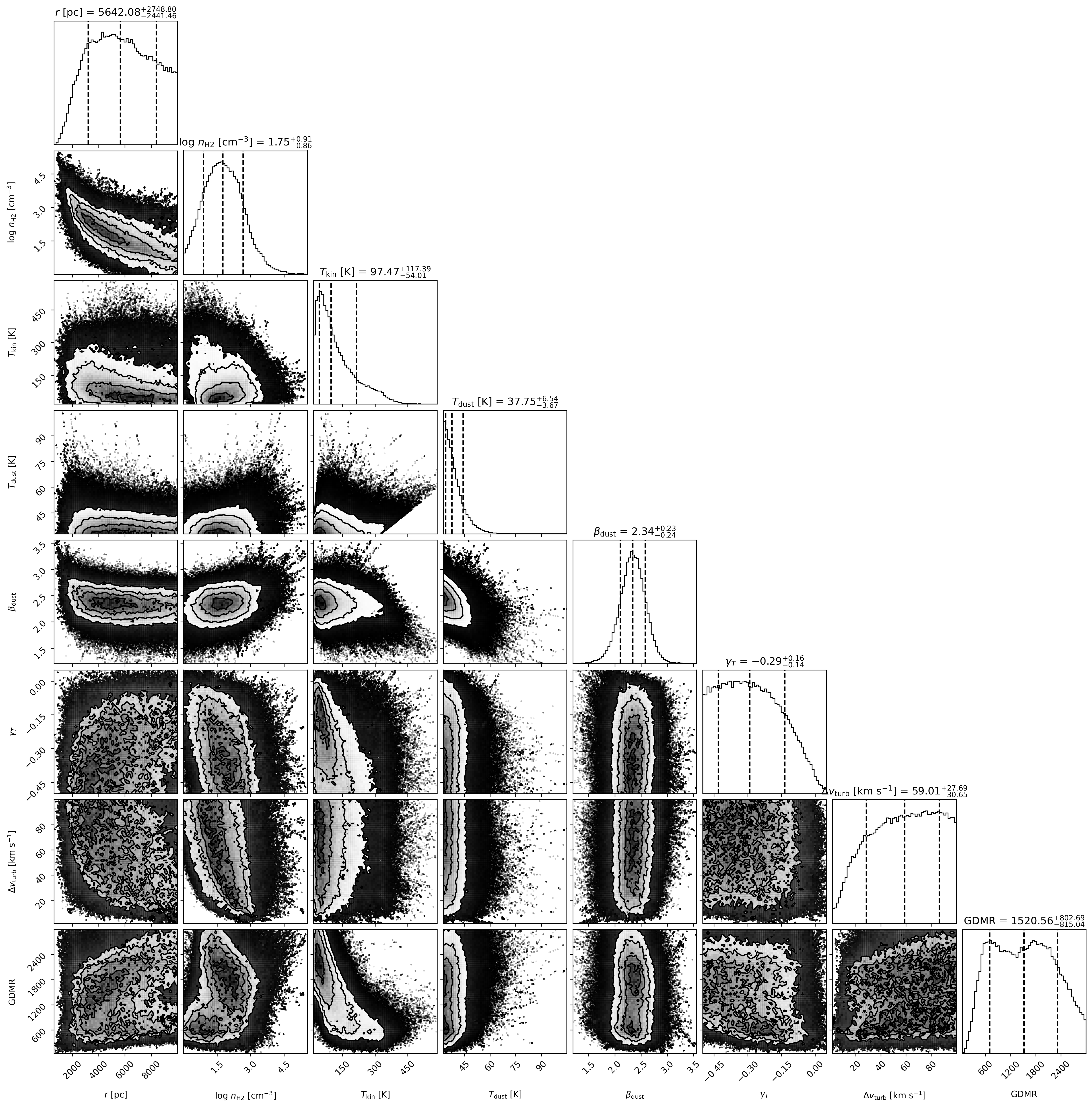}
 \caption{Corner plot \citep[][]{Foreman-Mackey_2016} of the posterior distributions of the main physical parameters inferred from the fiducial \texttt{TUNER} model for REBELS-25 (see Fig.~\ref{fig:tuner}). The lower panels show the 2D marginalized covariances, while the top panels show the 1D marginalized posteriors with the median, 16th- and 84th-percentiles.}
 \label{fig:tuner_appendix}
\end{figure*}

In this Appendix, we provide additional details about the setup, assumptions, and posterior distributions of the \texttt{TUNER} modelling used in Sec.~\ref{TUNER}. 
In Table~\ref{tab:tuner_fluxes} we report all continuum and line flux measurements of REBELS-25 used in the \texttt{TUNER} model. 
Figure~\ref{fig:tuner_appendix} shows the corner plot of the posteriors for the fiducial \texttt{TUNER} model shown in Fig.~\ref{fig:tuner}. 

In the modelling, we fix the virial parameter to $k_{\rm vir}=1$, assuming the molecular gas is gravitationally bound. We adopt uniform priors on all free parameters, chosen to span a wide but physically motivated range appropriate for a massive star-forming galaxy at $z>7$, and informed by the extensive ancillary data available for REBELS-25. 
The radius of the molecular gas reservoir is allowed to vary between $0.1$ and $10^4\,$pc. The mean molecular hydrogen density spans $\log(n_{\rm H_2}/{\rm cm^{-3}})=0-7$. The kinetic temperature of the gas is allowed in the range $T_{\rm kin}=10-600\,$K, and its dependence on density is described by the power-law slope $\gamma_T$, which is allowed to span $-0.5-0.05$.
The turbulent velocity width, which sets the width of the lognormal density distribution, is allowed to vary between $1$ and $100\,$\kms. The upper bound is slightly higher than what we infer translating the velocity dispersion derived from the [C\,\textsc{ii}] kinematic modelling from \citet{Rowland_2024} into a FWHM ($\Delta v_{\rm turb}\simeq78\,$\kms). 
We allow for a wide range of gas-to-dust mass ratios, $\delta_{\rm GDR}=1-3000$, reflecting the elevated and scattered values reported for other high-redshift sources \citep[see e.g.,][]{Algera_2026}. 
The prior on the dust temperature is coupled to the gas kinetic temperature as $0.5\leq T_{\rm kin}/T_{\rm dust} \leq 10$ and the dust emissivity slopes is allowed to vary over $\beta_{\rm dust}=1.2-3.6$.

For the dust opacity, we adopt the prescription of \citet{Weingartner_Draine_2001}, normalizing the opacity at rest-frame $158\,\mu$m with $\kappa_0=1.041\,$cm$^2\,$g$^{-1}$. This choice differs from the default \texttt{TUNER} implementation, which typically uses the value from \citet{Draine_2014} anchored at $850\,\mu$m. While the two prescriptions are consistent for the fiducial emissivity slope, anchoring the opacity at $158\,\mu$m is better matched to the wavelength range probed by the observed dust continuum of REBELS-25. For non-fiducial values of $\beta_{\rm dust}$, the anchoring wavelength can have a non-negligible impact on the inferred dust and gas masses.

One notable aspect of the \texttt{TUNER} model is that it sets a temperature floor above $T_{\rm CMB}(z)$, to ensure the temperature of the material stays slightly above the CMB, and thus preventing large amounts of gas from becoming effectively invisible due to a lack of contrast against the background. This floor can be adjusted in the model setup prior to each run. 
In its standard configuration, the model assumes a floor of $10\,$K, motivated by conditions in the local Universe where $T_{\rm CMB}=2.73\,$K; under this assumption, dust and gas temperatures are not allowed to fall below $\simeq13\,$K, consistent with typical cold ISM temperatures.
At the redshift of REBELS-25, however, a $10\,$K floor would impose a minimum temperature of $\simeq33\,$K, comparable to both the dust temperature inferred for this source \citep[][]{Algera_2024b} and the excitation temperature of the CO(3--2) transition. 
We therefore lower this temperature floor to $3\,$K to avoid biasing our results to higher temperatures. 

We also explored the impact of choosing different values for the temperature floor over $T_{\rm CMB}$. We run several models, allowing different temperature floors (from $0\,$K to $10\,$K)  to assess their impact on the inferred ISM properties. We find that when the floor is set to $\lesssim2\,$K, the models tend to favour nearly isothermal gas conditions and struggle to converge on physically meaningful solutions, accompanied by abrupt changes in the shapes of several posterior distributions. Conversely, adopting a floor of $\gtrsim7\,$K forces the dust temperature to remain above $\simeq30\,$K, in tension with the results from the MBB fits to the observed continuum emission.

\begin{table}
	\centering
    \caption{REBELS-25 continuum and line flux measurements used in the \texttt{TUNER} model. Non-detections are quoted as $1\sigma$ upper limits. Reference [1]: \citet{Algera_2024b}.}
	\label{tab:tuner_fluxes}
	\begin{tabular}{lcl} 
		\hline
            \hline
		Flux & Value &  Reference \\
		\hline
		$S_{41.3\,\rm GHz}$ (VLA Q Band) & $<5.3\,\mu$Jy & This work \\
        $S_{102.5\,\rm GHz}$ (ALMA Band 3) & $16.9\pm3.9\,\mu$Jy & [1] \\ 
        $S_{145.9\,\rm GHz}$ (ALMA Band 4) & $34.9\pm7.8\,\mu$Jy & [1] \\ 
        $S_{168.9\,\rm GHz}$ (ALMA Band 5) & $81.3\pm10.1\,\mu$Jy & [1] \\ 
        $S_{227.3\,\rm GHz}$ (ALMA Band 6) & $226.5\pm13.9\,\mu$Jy & [1] \\ 
        $S_{403.4\,\rm GHz}$ (ALMA Band 8) & $641.4\pm146.9\,\mu$Jy & [1] \\
        $S_{687.6\,\rm GHz}$ (ALMA Band 9) & $<366\,\mu$Jy & [1] \\
        \hline
        $S_{\rm CO(3-2)}\Delta v$ (VLA Q Band) & $88\pm21\,$mJy$\,$\kms & This work \\
        $S_{\rm CO(7-6)}\Delta v$ (ALMA Band 3) & $64\pm23\,$mJy$\,$\kms & This work \\ 
		\hline
	\end{tabular}
\end{table}

\bsp	
\label{lastpage}
 \end{document}